\documentclass[12pt,a4paper,final]{iopart}

\usepackage{iopams}
\usepackage[breaklinks=true,colorlinks=true,linkcolor=blue,urlcolor=blue,citecolor=blue]{hyperref}
\usepackage{amsfonts}

\expandafter\let\csname equation*\endcsname\relax
\expandafter\let\csname endequation*\endcsname\relax
\usepackage{amsmath}
\usepackage{amssymb}
\usepackage{bm}
\usepackage{dcolumn}
\usepackage{graphicx}
\usepackage{graphics}
\usepackage[latin1]{inputenc}
\usepackage{latexsym}
\usepackage{rotating}
\usepackage{hyperref}
\usepackage[all]{hypcap} 
\usepackage{xspace} % Sensible space treatment at end of simple macros
\usepackage[usenames,dvipsnames]{color}
\usepackage{mathrsfs}
\usepackage{subfigure}

\usepackage{ulem}
\usepackage{outlines}
\usepackage{enumitem}
\setenumerate[1]{label=\Roman*.}
\setenumerate[2]{label=\Alph*.}
\setenumerate[3]{label=\roman*.}
\setenumerate[4]{label=\alph*.}

\definecolor {darkgreen}{rgb}{0.2,0.7,0.2}

\newcommand\be{\begin{equation}}
\newcommand\ba{\begin{eqnarray}}
\newcommand\ee{\end{equation}}
\newcommand\ea{\end{eqnarray}}

\newcommand\bw{\begin{widetext}}
\newcommand\ew{\end{widetext}}

\newcommand{\GR}{{\mbox{\tiny GR}}}

\newcommand{\yn}[1]{\textcolor{blue}{\it{\textbf{yn: #1}}} }

\usepackage{here}

\begin{document}

\title[Weakly-Grav.~Objects in dCS gravity and constraints from GPB]
{Weakly-Gravitating Objects in dynamical Chern-Simons gravity and Constraints with Gravity Probe B}

\author{%
Yuya Nakamura$^{1}$,
Daiki Kikuchi$^{1}$,
Kei Yamada$^{2}$,
Hideki Asada$^{1}$
and
Nicol\'as~Yunes$^{3}$
}
\address{$^{1}$~Faculty of Science and Technology, Hirosaki University,
Hirosaki 036-8561, Japan.}
\address{$^{2}$~Department of Physics, Kyoto University, Kyoto 606-8502, Japan.}
\address{$^{3}$~eXtreme Gravity Institute, Department of Physics,
Montana State University, Bozeman, MT 59717, USA.}

\ead{k.yamada@tap.scphys.kyoto-u.ac.jp}

\date{\today}

%%%%%%%%%%%%%%%%%%%%%%%%%%%%%%%%%%%%%%%%%%%%%%%%%
\begin{abstract}
  
  Solar system observations have traditionally allowed for very stringent tests of Einstein's theory of general relativity. 
  We here revisit the possibility of using these observations to constrain gravitational parity violation as encapsulated in dynamical Chern-Simons gravity. 
  Working in the small-coupling and post-Newtonian approximations, we calculate analytically the scalar field and the gravitomagnetic sector of the gravitational field in the interior and the exterior of an isolated, weakly-gravitating object with uniform rotation and a quadrupolar mass deformation.
  We find that the asymptotic peeling-off behavior of the exterior fields is consistent with that found for black holes and neutron stars, as well as for non-relativistic objects, with overall coefficients that are different and dependent on the structure of the weak-field source. 
  We then use these fields to explicitly calculate the dynamical Chern-Simons correction to the spin precession of gyroscopes in orbit around Earth, which we find to be in the same direction as the Lense-Thirring effect of General Relativity.
  We then compare this correction to the spin precession prediction of General Relativity to the results of the Gravity Probe B experiment to place a constraint on dynamical Chern-Simons theory that is consistent with previous approximate estimates. 
Although we focus primarily on a single body, our methods can be straightforwardly extended to binary systems or N-bodies.
\end{abstract}

\maketitle
%\tableofcontents

%%%%%%%%%%%%%%%%%%%%%%%%%%%%%%%%%%%%%%%%%%%%%%%%%
\section{Introduction}

Is Einstein still right? Solar System observations~\cite{Will:2014kxa,2003Natur.425..374B} and binary pulsar observations~\cite{Stairs:2003eg} have confirmed that General Relativity (GR) is correct to extreme precision in the weak field regime, where the gravitational interaction is feeble and non-dynamical. The recent gravitational wave observations by the advanced Laser Interferometer Gravitational Wave Observatory~\cite{Abramovici:1992ah,Abbott:2007kv,Harry:2010zz,LIGOweb} and the advanced Virgo detector~\cite{Caron:1997hu,Giazotto:1988gw,TheVirgo:2014hva,Virgoweb} have confirmed that GR seems to still be on a firm footing, even in the \emph{extreme gravity} regime, where the gravitational interaction is not only strong, but also non-linear and highly dynamical~\cite{Berti:2018cxi,Berti:2018vdi}. 

Given these observations, one may then wonder whether further tests of GR are still necessary. The main arguments for continued testing are both observational and theoretical. On the observational side, the rotation curves of galaxies~\cite{1970ApJ...159..379R}, the late-time acceleration of the universe as inferred through supernova~\cite{Riess:2004nr}, and many other observations~\cite{1999A&A...350....1H,2006ApJ...648L.109C,PhysRevD.74.123507,0004-637X-633-2-560,2018arXiv180706209P} point to anomalies that can either be resolved by invoking the existence of dark matter~\cite{1994ARA&A..32..531C} and dark energy~\cite{PhysRevD.60.081301}, or alternatively, by modifying GR on large scales~\cite{PhysRevLett.93.011301}. On the theoretical side, the intrinsic incompatibility of GR with quantum mechanics have prompted the emergence of modifies theories that attempt a reconciliation~\cite{0264-9381-21-15-R01,2004quantumgrav.book.}. Regardless of the motivation, further testing may provide additional hints that may help in resolving some of these anomalies.

One particular modification to GR that has attracted some attention recently is gravitational parity-violation as encapsulated in dynamical Chern-Simons (dCS) gravity~\cite{Jackiw:2003pm,Alexander:2009tp}. This theory modifies Einstein's through a dynamical (pseudo) scalar field that couples non-minimally to curvature through the Pontryagin density.  The magnitude of any dCS deformation from GR is controlled by the size of its dimensional coupling parameter $\xi$. Today, we understand this theory as an \emph{effective} model, valid only at sufficiently low-energies/small-curvatures relative to some cut-off scale, or equivalently, for small coupling $\xi$. This is because the theory is motivated from heterotic string theory upon 4-dimensional compactification and a low-curvature expansion~\cite{Alexander:2004xd}, from loop quantum gravity upon the promotion of the Barbero-Immirzi parameter to a field in the presence of matter~\cite{Taveras:2008yf,Calcagni:2009xz}, and from effective field theories of inflation~\cite{Weinberg:2008hq}. 

Gravitational parity violation arises in this theory in the sense that GR deviations activate only for systems that violate parity through the presence of a preferred axis, like that represented by angular momentum in a dynamical system~\cite{Yunes:2008ua,Alexander:2007zg,Smith:2007jm,Alexander:2007vt,Yagi:2013mbt}. An example is an isolated spinning black hole, with dCS solutions known to fifth order in a slow-rotation expansion~\cite{Yunes:2009hc,Yagi:2012ya,Maselli:2017kic}, and in the near-extremal limit~\cite{McNees:2015srl,PhysRevD.97.084012}. Another example is a spinning star or planet, whose dynamics in the Solar System have been studied and observed for almost a century. One may thus expect that any dCS deviation from the GR predictions in the spin dynamics of the Solar System could be used to constrain the theory.   

Previous work has focused on this idea. In the context of an older version of this theory, the so-called non-dynamical version, Alexander and Yunes~\cite{Alexander:2007vt} showed that the gravitomagnetic sector of the metric (the temporal-spatial components) is modified in the Solar System, leading to a new parameterized post-Newtonian parameter and to a modification to Lense-Thirring precession~\cite{{Alexander:2007zg}}. In a similar study, Smith, et.~al.~\cite{Smith:2007jm} calculated this correction to spin-precession for a uniform density object rotating uniformly; they then compared the result to observations by the Gravity Probe B (GPB) experiment~\cite{Everitt:2011hp} and the LAGEOS satellites to place a stringent constraint on the non-dynamical theory. Shortly after, Ali-Ha{\"i}moud and Chen~\cite{AliHaimoud:2011fw} studied the dynamical theory, and among other calculations, they repeated parts of the study of Smith, et.~al.~\cite{Smith:2007jm}; in particular, they computed the correction to the gravitomagnetic sector of the metric for a uniform density object rotating uniformly to all orders in the coupling, and then used this metric correction to place an approximate constraint on the theory of $\xi_{\rm CS}^{1/4} \lesssim {\cal{O}}(10^{8} \; {\rm km})$. The possibility of a comparable constraint by quantum and/or Sagnac interferometry has also been discussed in~\cite{2012PhRvL.109w1101O,2013PhRvD..87h4038O,Kikuchi:2014mva}. In the future, the observation of gravitational waves emitted by spinning black holes will lead to an eight-order of magnitude improvement on these constraints~\cite{1806.07425v1}, once detections are made that are sufficiently strong to break degeneracies between the spins of the objects and the dCS deformation. 
	
In this paper, we revisit the study of observables in the dynamical theory analytically through the use of two approximations that are valid in the Solar System. First, we treat dCS gravity as an effective theory, thereby expanding all expressions in small coupling to guarantee the theory has a well-posed initial value problem~\cite{Delsate:2014hba}.  Second, we work in the post-Newtonian (PN) approximation, expanding all quantities assuming weak-fields and slow characteristic velocities~\cite{will1993theory,poisson2014gravity}. These approximations are sufficient to describe weakly-gravitating and slowly-rotating objects in dCS gravity in general. 

We then use these approximations to find analytic solutions for the scalar field and the gravitomagnetic components of the metric deformation in dCS gravity, both in the interior and the exterior of a weakly-gravitating body. We study both the case of a uniform density object, as well as the case of an object with a quadrupolar deformation, explicitly showing that the latter are greatly subdominant in the Solar System. In the case of uniform density, our solutions are similar to those that describe black holes and neutron stars in dCS gravity in the far-field~\cite{Yunes:2009hc,Yagi:2012ya,Yagi:2013mbt}, although of course our solutions depend on constants of integration that are specific to the matter distribution inside the gravitating object. Our uniform density solutions are also similar to those found for weakly-gravitating objects in~\cite{AliHaimoud:2011fw}, after expanding the latter in small coupling. Therefore, the solutions presented here are the first to self-consistently describe a weakly-gravitating and rotating object in post-Newtonian theory within the small-coupling approximation, and with or without a uniform density. 

With these solutions at hand, we then explicitly compute how gyroscopes would precess around Earth due to the spin angular momentum of our planet in dCS gravity. Unlike in the non-dynamical theory, we find that the dCS corrections to spin-precession are in the same direction as the effects that arise due to Lense-Thirring precession in GR. With these results at hand, we then compare the spin-precession rate of GPB to the dCS prediction, placing a constraint on the latter of $\xi_{\rm CS}^{1/4} \lesssim 1.3 \times 10^{8} \; {\rm km}$. This constraint is consistent with the order of magnitude estimate derived by Ali-Ha{\"i}moud and Chen by looking at the dCS metric perturbation~\cite{AliHaimoud:2011fw}.

The remainder of this paper is organized as follows. 
Section~\ref{Sec:Basics_dCS} briefly reviews the basics of dCS gravity.
Section~\ref{Sec:PN_Sols} presents the weak-field expansions we use to obtain solutions in dCS gravity.
Sections~\ref{Sec:const-density} and~\ref{Sec:quad} obtain analytic solutions for the interior and exterior scalar field and gravitomagnetic sector of the metric deformation for a weakly-gravitating and slowly-rotating body with uniform density and with a quadrupolar deformation respectively. 
Section~\ref{Sec:GPB} computes the dCS corrections to spin-precession for a gyroscope in orbit around Earth and compares these predictions with the observations of the GPB experiment. 
Section \ref{Sec:Conclution} concludes and points to future research. 
We adopt the conventions of~\cite{misner1973gravitation}, in particular for the signature of the metric, Riemann, and Einstein tensors. Greek letters represent spacetime indices, while Latin letters stand for spatial indices only. A semicolon and the $\nabla_{\mu}$ symbol denote the covariant derivatives compatible with the metric tensor $g_{\mu\nu}$, while a comma and the $\partial_{\mu}$ symbol stand for partial derivatives. Throughout this paper we use geometric units in which $G = 1 = c$.

%%%%%%%%%%%%%%%%%%%%%%%%%%%%%%%%%%%%%%%%%%%%%%%%%
\section{Basics of dynamical Chern-Simons modified gravity}
\label{Sec:Basics_dCS}

In this section, we describe the basics of dCS gravity. The action is given by~\cite{Alexander:2009tp}
\begin{align}
  \label{Action}
  S \equiv \int d^4 x \sqrt{-g} \left[ \kappa_g R + \frac{\alpha}{4} \vartheta \, {^\ast\!} R R - \frac{\beta}{2} \nabla_{\mu} \vartheta \nabla^{\mu} \vartheta + \mathcal{L}_{\rm mat} \right] ,
\end{align}
where $\kappa_g \equiv (16 \pi)^{- 1}$, $g$ denotes the determinant of the metric $g_{\mu \nu}$, $R$ is the Ricci scalar, $\mathcal{L}_{\rm mat}$ denotes the matter Lagrangian density, and $\alpha$ and $\beta$ are coupling constants. The Pontryagin density ${^\ast\!} R R$ is given by~\cite{Alexander:2009tp}
\begin{align}
  \label{Pont}
  {^\ast\!} R R \equiv {^\ast\!} R^{\mu \nu \rho \sigma} R_{\nu \mu \rho \sigma} , \qquad
  {^\ast\!} R^{\mu \nu \rho \sigma} \equiv \frac{1}{2} \epsilon^{\rho \sigma \alpha \beta} R^{\mu \nu}_{~~ \alpha \beta} ,
\end{align}
where $\epsilon^{\mu \nu \rho \sigma}$ is the Levi-Civita tensor. If the (pseudo) scalar field $\vartheta$ is constant, dCS gravity reduces identically to GR, because the Pontryagin density term in the action becomes the total divergence of a topological current~\cite{Jackiw:2003pm}, and therefore it does not contribute to the field equation.
We 
take $\vartheta$ and $\beta$ to be dimensionless, which forces $\alpha$ to have a dimension of $(\rm length)^2$.

The field equations of dCS gravity are obtained by varying the action [Eq.~\eqref{Action}] with respect to the metric $g_{\mu \nu}$ and the scalar field $\vartheta$~\cite{Alexander:2009tp}:
\begin{align}
  \label{eq;metric}
  G_{\mu \nu} + \frac{\alpha}{\kappa_g} C_{\mu \nu}
  &= \frac{1}{2 \kappa_g} \left( T^{\rm mat}_{\mu \nu} + T^{\vartheta}_{\mu \nu} \right) , \\
  \label{eq;scalar}
  \Box \vartheta
  &= - \frac{\alpha}{4 \beta} {^\ast\!} R R ,
\end{align}
where $G_{\mu \nu}$ is the Einstein tensor and $T^{\rm mat}_{\mu \nu}$ is the stress-energy tensor for the matter field.  The d'Alembertian operator is here denoted by $\Box \equiv \nabla_{\alpha} \nabla^{\alpha}$.
The C-tensor and the stress-energy tensor for the scalar field are defined by 
\begin{align}
  C^{\mu \nu} &\equiv \left( \nabla_{\sigma} \vartheta \right) \epsilon^{\sigma \delta \alpha ( \mu} \nabla_{\alpha} R^{\nu )}{}_{\delta} + \left( \nabla_{\sigma} \nabla_{\delta} \vartheta \right) {^\ast\!} R^{\delta ( \mu \nu ) \sigma} , \\
  T^{\vartheta}_{\mu \nu} &\equiv \beta \left( \nabla_{\mu} \vartheta \right) \left( \nabla_{\nu} \vartheta \right) - \frac{\beta}{2} g_{\mu \nu} \nabla_{\delta} \vartheta \nabla^{\delta} \vartheta ,
\end{align}
where the parenthesis in subscripts stand for symmetrization of indices. The equation of motion for the scalar field, Eq.~\eqref{eq;scalar}, may have a homogeneous solution, but this solution will render dCS gravity very similar to the non-dynamical theory, and thus, severely constrained by cosmological observations~\cite{Dyda:2012rj}. For the rest of this paper, we will ignore the homogeneous solution for $\vartheta$, and instead, concentrate on dynamically generated scalar fields.

Iterative solutions of Eqs.~\eqref{eq;metric} and~\eqref{eq;scalar} in a PN expansion are most easily achieved by recasting the field equations in trace-reversed form, namely 
\begin{align}
  \label{eq;trace-reverced}
  R_{\mu \nu} = - \frac{\alpha}{\kappa_g} C_{\mu \nu} + \frac{1}{2 \kappa_g} \left( \bar{T}^{\rm mat}_{\mu \nu} + \bar{T}^{\vartheta}_{\mu \nu} \right) , 
\end{align}
where we have used that the trace of C-tensor vanishes identically. The trace-reversed tensors $\bar{T}^{\rm mat}_{\mu \nu}$ and $\bar{T}^{\vartheta}_{\mu \nu}$ are simply 
\begin{align}
  \bar{T}^{\rm mat}_{\mu \nu} &\equiv T^{\rm mat}_{\mu \nu} - \frac12 g_{\mu \nu} T^{\rm mat} , \\
  \bar{T}^{\vartheta}_{\mu \nu} &\equiv \beta \left( \nabla_{\mu} \vartheta \right) \left( \nabla_{\nu} \vartheta \right) ,
\end{align}
where $T^{{\rm mat}} = g^{\mu \nu} T^{\rm mat}_{\mu \nu}$ is the trace of the matter stress-energy tensor. 

DCS gravity ought to be thought as an effective field theory, valid only as an expansion about weak curvatures and low energies. As such, one should really treat the theory as a deformation of GR, thus requiring a \emph{small-coupling} expansion of solutions of the form~\cite{Alexander:2007vt}
\begin{align}
  \label{eq:small-coup-g}
  g_{\mu \nu} &= g_{\mu \nu}^{\GR} + \delta h_{\mu \nu} , \\
  \label{eq:small-coup-th}
  \vartheta &= \zeta^{1/2} \vartheta^{\GR} ,
\end{align}
where $g_{\mu \nu}^{\GR}$ is a GR solution, $\delta h_{\mu \nu} = \mathcal{O} (\zeta)$ is a deformation to this solution and $\vartheta^{\GR}$ is the scalar field solution sourced by a GR solution, with $\zeta \ll 1$ a small dimensionless coupling parameter. The structure of the field equations reveals that 
\begin{align}
  \zeta \equiv \frac{\xi_{\rm CS}}{\mathcal{L}^4} , 
  \qquad
  \xi_{\rm CS} \equiv \frac{\alpha^2}{\kappa_g \beta} ,
  \label{Eq:xiCS}
\end{align} 
where $\xi^{1/4}_{\rm CS} = \alpha^{1/2}/(\kappa_{g} \beta)^{1/4}$ is the characteristic length scale of the theory, while $\mathcal{L}$ is the characteristic length scale of the curvature of the system under consideration. For example, when dealing with a bound matter source, this length scale could correspond to the curvature length 
of the object.
The small-coupling treatment of dCS gravity suggests one should think of the C-tensor $C_{\mu \nu}$ in Eq.~\eqref{eq;trace-reverced} as a source for the metric tensor, because it is determined by $\vartheta^{\GR}$, which in turn is sourced by the GR solution $g_{\mu \nu}^{\GR}$.

%%%%%%%%%%%%%%%%%%%%%%%%%%%%%%%%%%%%%%%%%%%%%%%%%
\section{Weak Field Expansions in dCS Gravity}
\label{Sec:PN_Sols}

In this section, we explain the expansions we will employ to iteratively solve Eqs.~\eqref{eq;scalar} and~\eqref{eq;trace-reverced} for a weakly-gravitating and slowly-rotating object. We use both the small-coupling and the PN approximations, which then require that we expand all fields in Eqs.~\eqref{eq:small-coup-g} and~\eqref{eq:small-coup-th} in $\varepsilon \ll 1$, where $\varepsilon \sim M/R$ is the PN expansion parameter. In addition, we assume the slow rotation of the object, i.e., $\chi \ll 1$, where $\chi \equiv J/M^2$ is the dimensionless spin parameter and the spin angular momentum is $|\bm{J}|=J$. Hereafter, we use $\alpha'$ and $\varepsilon'$ as book-keeping parameters to label the order kept in the small-coupling and the PN expansions of the solution, where for example any term linear in $\zeta \propto \alpha^2$ and/or linear in $\varepsilon$ will be said to be of ${\cal{O}}(\varepsilon',\alpha'^{2})$. 

We assume the matter field can be described by a perfect fluid, whose stress-energy tensor is given by 
\begin{align}
  T^{\rm mat}_{\mu \nu} = ( \rho + \rho \Pi + p ) u^{\mu} u^{\nu} + p g^{\mu \nu} ,
\end{align}
where $\rho$, $p$, and $\Pi$ are its mass, pressure, and specific energy density, while $u^{\mu}=d x^{\mu}/d \tau$ is the four-velocity of the fluid element~\cite{will1993theory}. Such a stress-energy tensor suffices to describe the metric tensor exterior to planets, such as Earth, and stars, such as the Sun, in the PN approximation. We further decompose the GR background around Minkowski spacetime $\eta_{\mu \nu}$, turning Eq.~\eqref{eq:small-coup-g} into
\begin{align}
  g_{\mu \nu} = \eta_{\mu \nu} + h_{\mu \nu} + \delta h_{\mu \nu} ,
  \label{Eq:MetPert}
\end{align}
where $h_{\mu \nu}$ is the PN metric perturbation in GR. Hereafter, we work in Cartesian coordinates in which the Minkowski metric takes the simple form $\eta_{\mu \nu} = {\rm diag} (-1, 1, 1, 1)$, and we raise and lower indices with $\eta_{\mu \nu}$, as we work to leading order in the small-coupling and PN approximations. 

Before continuing, let us briefly summarize the PN expansion of the GR solution $h_{\mu \nu}$ following~\cite{will1993theory}, so as to establish some useful notation for future sections. Substituting Eq.~\eqref{Eq:MetPert} into Eq.~\eqref{eq;trace-reverced} in the GR limit (i.e.~$\alpha \to 0$, which implies $\delta h_{\mu \nu} \to 0$ and $\vartheta \to 0$), and imposing the standard PN gauge conditions
\begin{align}
  \label{PNgauge1}
  h^{~ k}_{j ~ , k} - \dfrac{1}{2} h_{, j} = \mathcal{O} ( \varepsilon'^{\, 4} ) , \\
  \label{PNgauge2}
  h^{~ k}_{0 ~ , k} - \dfrac{1}{2} h^{k}_{~ k , 0} = \mathcal{O} ( \varepsilon'^{\, 5} ) , 
\end{align}
where $h = \eta^{\mu \nu} h_{\mu \nu}$ and $h^{i}_{~ i} = \delta^{i j} h_{i j}$, the PN expansion of the GR solution is 
\begin{align}
  \label{1PN}
  h_{0 0} &= 2 U - 2 U^2 + 4 \Phi_1 + 4 \Phi_2 + 2 \Phi_3 + 6 \Phi_4 + \mathcal{O} ( \varepsilon'^{\, 6} ) , \nonumber \\
  h_{0 i} &= - \frac72 V_i - \frac12 W_i + \mathcal{O}( \varepsilon'^5 ) , \nonumber \\
  h_{i j} &= 2 U \delta_{i j} + \mathcal{O}( \varepsilon'^4 ) .
\end{align} 
The PN potentials $U, \Phi_1, \Phi_2, \Phi_3, \Phi_4, V_i, W_i$ are defined by
\begin{alignat}{5}
  U &\equiv \int \frac{\rho( {\bm{x}}' )}{|{\bm{x}} - {\bm{x}}'|} d^3 x' , \nonumber \\
  V_i &\equiv \int \frac{\rho( {\bm{x}}' ) v_i( {\bm{x}}' )}{|{\bm{x}} - {\bm{x}}'|} d^3 x' , \nonumber \\
  W_i &\equiv \int \frac{\rho( {\bm{x}}' ) v_j( {\bm{x}}' ) ( x - x' )^j ( x - x' )_i}{|{\bm{x}} - {\bm{x}}'|^3} d^3 x' , \nonumber \\
  \Phi_1 &\equiv \int \frac{\rho( {\bm{x}}' ) v^2( {\bm{x}}' )}{|{\bm{x}} - {\bm{x}}'|} d^3 x' ,\nonumber \\
  \Phi_2 &\equiv \int \frac{\rho( {\bm{x}}' ) U( {\bm{x}}' )}{|{\bm{x}} - {\bm{x}}'|} d^3 x' , \nonumber \\
  \Phi_3 &\equiv \int \frac{\rho( {\bm{x}}' ) \Pi( {\bm{x}}' )}{|{\bm{x}} - {\bm{x}}'|} d^3 x' ,\nonumber \\
  \Phi_4 &\equiv \int \frac{p( {\bm{x}}' )}{|{\bm{x}} - {\bm{x}}'|} d^3 x' ,
\end{alignat}
where the three-vector ${\bm x}$ denotes position of the field point and $v^i = u^i/u^0$. These PN potentials satisfy the following Poisson equations
\begin{align}
  \nabla^2 U &= - 4 \pi \rho , \nonumber \\
  \nabla^2 V_i &= - 4 \pi \rho v_i , \nonumber \\
  \nabla^2 \Phi_1 &= - 4 \pi \rho v^2 , \nonumber \\
  \nabla^2 \Phi_2 &= - 4 \pi \rho U , \nonumber \\
  \nabla^2 \Phi_3 &= - 4 \pi \rho \Pi , \nonumber \\
  \nabla^2 \Phi_4 &= - 4 \pi p ,
\label{eq:PN-Pot}
\end{align}
where $\nabla^2 = \eta^{i j} \partial_i \partial_j$ is the Laplacian operator in flat space. The Poisson equations presented above can be solved once the properties of the matter distribution are chosen; in what follows, we will solve these equations, together with their dCS deformations, for a constant density star, as well as a quadrupolarly deformed star. 

%%%%%%%%%%%%%%%%%%%%%%%%%%%%%%%%%%%%%%%%%%%%%%%%%
\section{Weak Field Solution in dCS Gravity: Constant Density}
\label{Sec:const-density}

In this section, we iteratively solve Eqs.~\eqref{eq;scalar} and~\eqref{eq;trace-reverced} for a weakly-gravitating and slowly-rotating object of constant density. We begin by solving the Poisson equations of the previous section for such an object, and then, using these solutions as the source for the PN expanded evolution equation for the scalar field. After solving for the evolution of this field, we then solve for the dCS metric deformation in the gravitomagnetic sector, which is sufficient to calculate the dCS modification to the GPB experiment. 

%----------------------------------------------------------------------------------------------------
\subsection{PN Solution for the Scalar Field}

Let us now consider the evolution equation for the scalar field [Eq.~\eqref{eq;scalar}]. Since the scalar field is linear in $\alpha$ in the small-coupling approximation, we must evaluate the Pontryagin density to zeroth-order in $\alpha$. This density at leading PN order is given by
\begin{align}
  \label{RR-PN}
  {^\ast\!} R R
  &= 4 \tilde{\epsilon}^{0 i j k} \left[h_{0 0 , i}^{~~~~ l} h_{j [l , 0] k} + h_{j ~ , k}^{~[l~\, m]} \left( h_{0 [ l , m ] i} - h_{i [ l , m ] 0} \right) \right] + \mathcal{O} (\varepsilon'^{\,6}, \alpha') \nonumber \\
  &= - 32 \tilde{\epsilon}^{0 i j k} \left( \partial_i \partial_l U \right) \left( \partial^l \partial_j V_k \right) + \mathcal{O}( \varepsilon'^{\,6}, \alpha' ) , 
\end{align}
where $\tilde{\epsilon}^{\alpha \beta \gamma \delta}$ is the Levi-Civita symbol with convention $\tilde{\epsilon}^{0123} = - \tilde{\epsilon}_{0123} = 1$, and the square brackets in subscripts stand for anti-symmetrization. In the last equality, we have substituted Eq.~\eqref{1PN} and used the fact $\tilde{\epsilon}^{0 i j k} \partial_j V_k = \tilde{\epsilon}^{0 i j k} \partial_j W_k$. Substituting this expression into Eq.~\eqref{eq;scalar}, we obtain
\begin{align}
  \nabla^2 \vartheta = \frac{8 \alpha}{\beta} \tilde{\epsilon}^{0 i j k} \left( \partial_i \partial_l U \right) \left( \partial^l \partial_j V_k \right) + \mathcal{O}( \varepsilon'^{\, 6}, \alpha'^{\, 2} ) 
\end{align}
where we have neglected the time derivatives of the scalar field on the left-hand side because they are of higher PN order; these expressions agree with Eq. (52) in~\cite{Yagi:2013mbt}. 

Let us assume that the object of interest has a uniform mass density $\rho = {\rm const.}$ and a constant angular velocity along the $z$-axis $\omega^{i} = (0, 0, \omega)$. The PN potentials $U$ and $V_k$ then become
\begin{align}
  \label{Eq:U_1PN}
  & U =
    \begin{cases}
      \dfrac{M}{r} & ( r > R ), \\
      2 \pi \rho \left( R^2 - \dfrac{r^2}{3} \right) & ( r \leq R ) , 
    \end{cases}
  \\
  \label{Eq:V_1PN}
  & V_k =
    \begin{cases}
      \vspace{0.2cm}
      \dfrac{1}{5} M R^2 \, \tilde{\epsilon}^0_{~ i j k} \, \omega^i \dfrac{x^j}{r^3} & ( r > R ) , \\ 
      \dfrac{2 \pi}{3} \rho \left( R^2 - \dfrac35 r^2 \right) \tilde{\epsilon}^0_{~ i j k} \, \omega^i x^j  & ( r \leq R ) , 
    \end{cases}
\end{align}
where $r$ denotes the distance from the center of the object to the field point, $M$ is the mass of the object and $R$ is its radius. Substituting these expressions into Eq.~\eqref{RR-PN}, we obtain 
\begin{align}
  \label{RRPN}
  {^\ast\!} R R = 
  \begin{cases}
    288 \dfrac{M^3 \chi}{r^7} \cos \theta + \mathcal{O} ( \varepsilon'^{\, 6}, \alpha' ) & ( r > R ) , \\
    \mathcal{O}( \varepsilon'^{\, 6}, \alpha' ) & ( r \leq R ) , 
  \end{cases}
\end{align}
where $\theta$ is the polar angle in polar coordinates, related to the $z$ coordinate via the standard relation $z = r \cos{\theta}$, while $\chi = I \omega/M^2$ with $I$ the moment of inertia, and we have used that
$I = 2 M R^2 / 5$ for a solid sphere of constant density. 

The evolution equation for the scalar field then becomes 
\begin{align}
\label{scalarextint}
  \nabla^2 \vartheta =
  \begin{cases}
    - \dfrac{72 \alpha}{\beta} \dfrac{M^3 \chi}{r^7} \cos \theta + \mathcal{O} ( \varepsilon'^{\, 6}, \alpha'^{\, 2} )  & ( r > R ), \\
    \mathcal{O} ( \varepsilon'^{\, 6}, \alpha'^{\, 2} ) & ( r \leq R )  , 
  \end{cases}
\end{align}
where again we have neglected time derivatives on the left-hand side. Equation~\eqref{scalarextint} implies that $\vartheta \sim \varepsilon'^{\, 5} \alpha'$, and thus, dCS modifications to the metric tensor will be necessarily small. We should here also note that Eq.~\eqref{scalarextint} is source-free in the interior region $r \leq R$, because we have focused on a constant mass density object. For a general mass density distribution, this will not be the case and the interior source term will not vanish at this order~\cite{AliHaimoud:2011fw}. While such a term affects the exterior solution through the matching of interior and exterior solutions at the surface of the object, this effect is small, as we will show in the next section. For now, let us continue to consider a constant density object. 

In order to solve Eq.~\eqref{scalarextint}, we decompose $\vartheta(r, \theta)$ in terms of the Legendre polynomials as 
\begin{align}
  \label{eq:Leg-Dec-theta}
  \vartheta( r, \theta ) = \sum_{\ell = 0}^{\infty} \vartheta_{\ell}( r ) P_{\ell}( \cos \theta ) .
\end{align}
The source term in Eq.~\eqref{scalarextint} vanishes at leading PN order for all $\ell \neq 1$ modes. Similar results have been reported for black holes~\cite{Yunes:2009hc, Yagi:2012ya} and neutron stars~\cite{Yagi:2013mbt}, as well as for non-relativistic objects~\cite{AliHaimoud:2011fw}. Therefore, we focus on the $\ell = 1$ mode only, for which Eq.~\eqref{scalarextint} becomes
\begin{align}
  \label{eq;S-PN}
  \left[ \frac{1}{r^2} \frac{d}{d r} \left( r^2 \frac{\partial}{\partial r} \right) - \frac{2}{r^2} \right] \vartheta_1 =
  \begin{cases}
    - \dfrac{72 \alpha}{\beta}\dfrac{M^3 \chi}{r^7} + \mathcal{O}( \varepsilon'^{\, 6}, \alpha'^{\, 2} ) & ( r > R ) , \\
    \mathcal{O}( \varepsilon'^{\, 6}, \alpha'^{\, 2} ) & ( r \leq R ) . 
  \end{cases}
\end{align} 
The general interior and exterior solutions are
\begin{align}
  \vartheta_{\rm ext} &= \left( A_1 r + \dfrac{A_2}{r^2} - \dfrac{4 \alpha}{\beta} \dfrac{M^3 \chi}{r^5} \right) \cos \theta + \mathcal{O} (\varepsilon'^{\, 6}, \alpha'^{\, 2} ) , \\
  \vartheta_{\rm int} &= \left( B_1 r + \dfrac{B_2}{r^2} \right) \cos \theta + \mathcal{O} (\varepsilon'^{\, 6}, \alpha'^{\, 2} ) , 
\end{align}
where $A_i$ and $B_i$ ($i = 1,2$) are constants of integration. 

These constants of integration can be determined by requiring regularity and smoothness of the solution at its boundaries. Requiring that $\vartheta$ be finite as $r \to 0$ and $r \to \infty$ forces $A_1 = 0$ and $B_2 = 0$. The remaining two constants are determined by requiring continuity and differentiability at the surface of the object, namely, 
\begin{align}
  \lim_{\lambda \to 0^{+}} \left[ \vartheta_{\rm ext} ( R + \lambda ) - \vartheta_{\rm int} ( R - \lambda ) \right] = 0 , & \\
  \lim_{\lambda \to 0^{+}} \left[ \left. \frac{\partial \vartheta_{\rm ext}}{\partial r} \right|_{r = R + \lambda} - \left. \frac{\partial \vartheta_{\rm int}}{\partial r} \right|_{r = R - \lambda} \right] = 0 , &
\end{align}
The latter condition can be derived by integrating Eq.~\eqref{eq;S-PN}:
\begin{align}
  \lim_{\lambda \to 0^{+}} \left[ \left. \frac{\partial \vartheta_{\rm ext}}{\partial r} \right|_{r = R + \lambda} - \left. \frac{\partial \vartheta_{\rm int}}{\partial r} \right|_{r = R - \lambda} \right]
  &= \lim_{\lambda \to 0^{+}} \int^{R + \lambda}_{R - \lambda} \frac{d^2 \vartheta}{d r^2} d r \nonumber \\
  &= - 2 \lim_{\lambda \to 0^{+}} \int^{R + \lambda}_{R - \lambda} \left[ \frac{d}{d r} \left( \frac{\vartheta}{r} \right) + \frac{36 \alpha}{\beta} \frac{M^3 \chi}{r^7} \right] d r = 0 ,
\end{align}
at leading PN order. 

Imposing the above matching conditions, we obtain the axisymmetric dCS scalar field for a weakly-gravitating and slowly-rotating object in the small-coupling and in the PN approximations:
\begin{align}
  \label{Eq:theta_ext}
  \vartheta_{\rm ext} &= 8 \left(\frac{\alpha}{\beta M^{2}}\right) \left(\frac{M}{R}\right)^{3} \left(\frac{M}{r}\right)^{3} \left(\frac{z}{M}\right) \chi \left[ 1 - \frac{1}{2} \left( \frac{R}{r} \right)^3 \right] + \mathcal{O} ( \varepsilon'^{\, 6}, \alpha'^{\, 2} ) , \\
  \label{Eq:theta_int}
  \vartheta_{\rm int} &= 4 \left(\frac{\alpha}{\beta M^{2}}\right) \left(\frac{M}{R}\right)^{6} \left(\frac{z}{M}\right) \chi + \mathcal{O} ( \varepsilon'^{\, 6}, \alpha'^{\, 2} ). 
\end{align}
Notice that both $ \vartheta_{\rm ext}$ and $\vartheta_{\rm int}$ are proportional to $z$, and thus to $\cos\theta$, which is consistent with field being a pseudo scalar. The exterior solution $\vartheta_{\rm ext}$ behaves as $1/r^2$ as $r \to \infty$, which agrees with the weak-field expansion for black hole~\cite{Yunes:2009hc,Yagi:2012ya} and neutron star solutions~\cite{Yagi:2013mbt}.

%----------------------------------------------------------------------------------------------------
\subsection{PN Solution for the dCS Metric Deformation: Gravitomagnetic Sector}

Let us now consider the field equations for the dCS metric deformation.
Since the stress-energy tensor for the matter field $T^{\rm mat}_{\mu \nu}$ is balanced by the GR solution, this does not source the dCS correction $\delta h_{\mu \nu}$ at the leading order. The field equations to first order in the small-coupling approximation and to leading-order in the PN approximation are then
\begin{align}
  \label{Eq:LMFE}
  \delta R_{\mu \nu} [\delta h] = - \frac{\alpha}{\kappa_g} C_{\mu \nu} [h, \vartheta] + \bar{T}^{\vartheta}_{\mu \nu} ,
\end{align}
where $\delta R_{\mu \nu}$ is the linearized Ricci tensor evaluated on the dCS metric deformation $\delta h_{\mu \nu}$.

The different pieces of the modified field equations can be easily calculated in the PN approximation. The C-tensor in the right-hand side of Eq.~\eqref{Eq:LMFE} can be expanded as
\begin{align}
  C_{0 0} &= - \frac12 \vartheta_{, i} \, \tilde{\epsilon}^{0 i j k} \nabla^2 h_{0 k, j} + \vartheta_{, i}^{~\, l} \, \tilde{\epsilon}^{0 i j k} h_{k [l, 0] j} + \mathcal{O}( \varepsilon'^{\, 9}, \alpha'^{\, 3} ) ,  \\
  C_{0 i} &= \frac12 \left[ \tilde{\epsilon}^{0 l j k} ( \vartheta_{, l} h_{ k [ m , i ] j} )^{, m} -  \frac12 \tilde{\epsilon}^{0 ~ j k}_{~ i} ( \vartheta_{, j} h_{0 0, k l} )^{, l} \right] + \mathcal{O} ( \varepsilon'^{\, 8}, \alpha'^{\, 3} ) ,  \\
  C_{i j} & = \tilde{\epsilon}^{0 m n}_{~~~~\, (i}  \left[ \vartheta_{, m} \left( \nabla^2 h_{n [ j ), 0 ]} + h_{l [ 0 , n ] j )}^{~~~~~~~ l} - \frac{1}{2} h_{0 0 , j) n 0} \right) + \vartheta_{, m}^{~~\, l} \left( h_{n [ j ), l ] 0} - h_{0 [ j ), l ] n} \right) \right] + \mathcal{O}( \varepsilon'^{\, 9}, \alpha'^{\, 3} )\,,
\end{align}
where we assume that the scalar field is stationary for the last equation, although this is not relevant below.
The stress-energy tensor of the scalar field $\bar{T}^{\vartheta}_{\mu \nu} = \mathcal{O} ( \varepsilon'^{\, 10}, \alpha'^{\, 2} )$, which is higher order in the PN approximation than what we consider here, so it can be safely neglected.
The linearized Ricci tensor is simply
\begin{align}
  \label{Eq:deltaR00}
  \delta R_{0 0} &= - \frac12 \nabla^2 \delta h_{0 0} + \mathcal{O} ( \varepsilon'^{\, 9}, \alpha'^{\, 3} ) , \\
  \delta R_{0 i} &= - \frac12 \nabla^2 \delta h_{0 i} + \mathcal{O} ( \varepsilon'^{\, 8}, \alpha'^{\, 3} ) , \\
  \label{Eq:deltaRij} 
  \delta R_{i j} &= - \frac12 \nabla^2 \delta h_{i j} - \delta h_{0 ( i, j ) 0} + \mathcal{O}( \varepsilon'^{\, 9}, \alpha'^{\, 3} ) , 
\end{align}
where we have imposed the standard PN gauge conditions on the dCS metric perturbation, namely
\begin{align}
  \label{Eq:ModPNgauge1} 
  \delta h^{~ k}_{j ~ , k} - \frac12 \delta h_{, j} &= \mathcal{O} ( \varepsilon'^{\, 9}, \alpha'^{\, 3} ) , \\
  \delta h^{~ k}_{0 ~ , k} - \frac12 \delta h^{k}_{~ k , 0} &= \mathcal{O} ( \varepsilon'^{\, 8}, \alpha'^{\, 3} ) . 
  \label{Eq:ModPNgauge2}
\end{align}

%----------------------------------------------------------------------------------------------------
Let us now focus on the gravitomagnetic part of Eq.~\eqref{Eq:LMFE}, namely
\begin{align}
  \label{gramag0}
  \nabla^2 \delta h_{0 i} = \frac{2 \alpha}{\kappa_g} \tilde{\epsilon}_{0 i j k} \left( \vartheta^{, j} \, \nabla^2 U^{, k} + \vartheta_{, \, j}^{~~ l} U_{, \,l}^{~~ k} \right) + \mathcal{O} (\varepsilon'^{\, 9}, \alpha'^{\, 3} ) .
\end{align}
Substituting Eqs.~\eqref{1PN},~\eqref{Eq:U_1PN},~\eqref{Eq:theta_ext}, and~\eqref{Eq:theta_int} into Eq.~\eqref{gramag0}, we obtain 
\begin{align}
  \label{eq;H}
  \nabla^2 H =
  \begin{cases}
    144 \zeta \dfrac{M^2 R^3 \chi}{r^7} \left( 1 - \frac{R^3}{r^3} \right) \sin \theta \, e^{- i \phi} + \mathcal{O} ( \varepsilon'^{\, 8}, \alpha'^{\, 3} ) & ( r > R ) , \\
    \mathcal{O} ( \varepsilon'^{\, 8}, \alpha'^{\, 3} )  & ( r \leq R ) ,
  \end{cases}
\end{align}
where we have defined $H \equiv \delta h_{0 y} + i \delta h_{0 x}$ with $i \equiv \sqrt{-1}$ and $\phi$ the azimuth in spherical polar coordinates. Henceforth, the dCS deformation parameter
\begin{align}
  \label{eq:zeta-def}
  \zeta = \xi_{\rm CS} \left( \frac{M}{R^3} \right)^2 ,
\end{align}
since the curvature scale of interest to our problem is ${\cal{L}} = M^{1/2}/R^{3/2}$. 

We can exploit the axisymmetry of our problem to immediately write down the solution along the rotation axis.
The gravitomagnetic component $\delta h_{0 z}$ must satisfy $\nabla^2 \delta h_{0 z} = 0$ over the entire spacetime because the rotation axis of the object is along the $\hat{z}$-direction.
The general solution to this equation is
\begin{align}
  \delta h_{0 z} = \sum_{\ell = 0}^{\infty} \left( C^{z}_{\ell} r^{\ell + 1} + D^{z}_{\ell} R^{- \ell} \right) P_{\ell}( \cos \theta ) ,
\end{align}
where $C^{z}_{\ell}$ and $D^{z}_{\ell}$ are constants of integration. By assuming regularity at the center of the object and asymptotic flatness at spatial infinity, we again find that $C^{z}_{\ell} = 0 = D^{z}_{\ell}$ for all $\ell$, and hence, $\delta h_{0 z} = 0$. As in the case of the scalar field, while Eq.~\eqref{eq;H} in the interior region $r \leq R$ is again source-free for a constant mass density distribution, a more general distribution would not be; the differences induced in the general solution will be small for the purpose of constraining the theory with weak-field, Solar System experiments, as we will show in the next section.

Decomposing $H$ into spherical harmonics $Y_{\ell}^m( \theta, \phi )$,
\begin{align}
  \label{eq:H-decomp}
  H ( r, \theta, \phi ) = \sum_{\ell, m} H_{\ell}^m( r ) Y_{\ell}^m( \theta, \phi ) ,
\end{align}
one finds that Eq.~\eqref{eq;H} is a homogeneous equation for all $\ell$-modes except for the $\ell = 1$ and $m = -1$ mode. Once more, this is in agreement with what one finds for black hole and neutron star solutions~\cite{Yunes:2009hc,Yagi:2012ya,Yagi:2013mbt}, as well as for non-relativistic objects~\cite{AliHaimoud:2011fw}. Since $H_{\ell}^m$ vanishes when $\ell \neq 1$ or $m \neq - 1$ by imposing asymptotic flatness and regularity at the center of the object, we focus on the $\ell = 1$ and $m = -1$ mode, which must satisfy 
\begin{align}
  \label{eq;H1-1}
  \frac{1}{r^2} \frac{d}{d r} \left( r^2 \frac{d H_1^{- 1}}{d r} \right) - 2 \frac{H_1^{- 1}}{r^2} =
  \begin{cases}
    144 \sqrt{\dfrac{8 \pi}{3}} \zeta \chi \dfrac{M^2 R^3}{r^7} \left( 1 - \frac{R^3}{r^3} \right) + \mathcal{O} ( \varepsilon'^{\, 8}, \alpha'^{\, 3} ) & ( r > R ) , \\
    \mathcal{O} ( \varepsilon'^{\, 8}, \alpha'^{\, 3} ) & ( r \leq R ) .
  \end{cases}
\end{align}
The general exterior and interior solutions are then
\begin{align}
  H_{\rm ext} &= \left[ C_1 r + \frac{C_2}{r^2} + 8 \sqrt{\frac{8 \pi}{3}} \zeta \chi \frac{M^2 R^3}{r^5} \left( 1 -  \frac13 \frac{R^3}{r^3} \right) \right] Y_1^{-1} ( \theta, \phi ) + \mathcal{O} ( \varepsilon'^{\, 8}, \alpha'^{\, 3} ) ,   \\
  H_{\rm int} &= \left( D_1 r + \frac{D_2}{r^2} \right) Y_1^{-1} ( \theta, \phi )+ \mathcal{O} ( \varepsilon'^{\, 8}, \alpha'^{\, 3} ) ,  
\end{align}
where $C_i$ and $D_i$ ($i = 1,2$) are integration constants. 

As before these constants can be determined by requiring regularity and smoothness at all boundaries. Asymptotic flatness and regularity at the center imply that $C_1 = 0$ and $D_2 = 0$, respectively, and thus, we need two more boundary conditions at the surface to determine the remaining constants. Requiring continuity at the surface, we have
\begin{align}
  \label{continuity}
  \lim_{\lambda \to 0^{+}} \left[ H_{\rm ext} ( R + \lambda ) - H_{\rm int} ( R - \lambda ) \right] = 0,
\end{align}
while requiring differentiability at the surface, we find 
\begin{align}
  \label{jump}
  \lim_{\lambda \to 0^{+}} \left[ \left. \frac{d H_{\rm ext}}{d r} \right|_{r = R + \lambda} - \left. \frac{d H_{\rm int}}{d r} \right|_{r = R - \lambda} \right]
  &= \lim_{\lambda \to 0^{+}} \int_{R - \lambda}^{R + \lambda} \frac{d^2 H}{d r^2} d r \nonumber \\
  &= - \frac{8 \pi \alpha}{\kappa_g} e^{- i \phi} \lim_{\lambda \to 0^{+}} \int_{R - \lambda}^{R + \lambda} \frac{1}{r} \frac{d \rho}{d r} \frac{\partial \vartheta}{\partial \theta} d r \nonumber \\
  &= - 24 \zeta \frac{M^2 \chi}{R^3} \sin \theta \, e^{- i \phi} ,
\end{align}
at leading PN order.
This is different from what we found in the scalar field case, since the condition above is a jump condition induced because Eq.~\eqref{gramag0} depends on a spatial derivative of the matter density.

With these boundary conditions at hand, we can now solve for the gravitomagnetic components of the dCS metric deformation. Imposing the boundary conditions in Eqs.~\eqref{continuity}-\eqref{jump}, one finds  
\begin{align}
  H_{\rm ext} &= 8 \, \zeta \, \left(\frac{M}{r}\right)^{2} \left(\frac{R}{r}\right)^3 \chi \left( 1 - \frac13 \frac{R^3}{r^3} \right) \sin \theta  \, e^{- i \phi} + \mathcal{O} ( \varepsilon'^{\, 8}, \alpha'^{\, 3} ) , \\
  H_{\rm int} &= \frac{16}{3} \, \zeta \, \left(\frac{M}{R}\right)^{2} \left(\frac{r}{R}\right) \chi \sin \theta \, e^{-i \phi} + \mathcal{O} ( \varepsilon'^{\, 8}, \alpha'^{\, 3} ) ,
\end{align} 
and using the definition of $H$ we then find
\begin{align}
  \label{Eq:delta_h0x_ext}
  \delta h^{\rm ext}_{0 x} &= - 8 \, \zeta \, \chi \left( \frac{M}{r} \right)^2 \left( \frac{R}{r} \right)^3 \left( \frac{y}{r} \right) \left( 1 - \frac13 \frac{R^3}{r^3} \right) + \mathcal{O} ( \varepsilon'^{\, 8}, \alpha'^{\, 3} ) , \\
  \delta h^{\rm int}_{0 x} &= - \, \frac{16}{3}  \, \zeta \, \chi \left( \frac{M}{R} \right)^2 \left( \frac{r}{R} \right) \left( \frac{y}{r} \right) + \mathcal{O} ( \varepsilon'^{\, 8}, \alpha'^{\, 3} ) , \\
  \label{Eq:delta_h0y_ext}
  \delta h^{\rm ext}_{0 y} &= 8 \, \zeta \, \chi \left( \frac{M}{r} \right)^2 \left( \frac{R}{r} \right)^3 \left( \frac{x}{r} \right) \left( 1 - \frac13 \frac{R^3}{r^3} \right) + \mathcal{O} ( \varepsilon'^{\, 8}, \alpha'^{\, 3} ) , \\
  \delta h^{\rm int}_{0 y} &= \frac{16}{3} \, \zeta \, \chi \left( \frac{M}{R} \right)^2 \left( \frac{r}{R} \right) \left( \frac{x}{r} \right) + \mathcal{O} ( \varepsilon'^{\, 8}, \alpha'^{\, 3} ) . 
\end{align}
The asymptotic peeling-off behavior of the exterior solutions agrees with the results of~\cite{Yunes:2009hc,Yagi:2012ya,Yagi:2013mbt,AliHaimoud:2011fw} in the weak-field limit as $r \to \infty$, while the overall coefficients are different and dependent on the structure of the weak-field source. This is reasonable in the small-coupling approximation because the scalar field should be determined by the background GR spacetime.

%%%%%%%%%%%%%%%%%%%%%%%%%%%%%%%%%%%%%%%%%%%%%%%%%
\section{Weak Field Solution in dCS Gravity: Quadrupolar Deformation}
\label{Sec:quad}

In this section, we repeat the calculation of Sec.~\ref{Sec:const-density} but for a weakly-gravitating source with a quadrupolar deformation, induced for example due to perturbations from an exterior body. As before, we begin by solving for the dCS scalar field and we then back-react this field onto the modified field equations to find the dCS metric deformation. We present the main skeleton of the calculation in this section, and defer details to Appendix~\ref{App:FullCal}. We focus once more on the gravitomagnetic sector of the metric, as this leads to the dominant dCS modification to the GPB experiment.  

%----------------------------------------------------------------------------------------------------
\subsection{PN Solution for the Scalar Field}

Consider a weakly-gravitating body with a non-constant mass density $\rho$ and a constant angular velocity along the $\hat{z}$-axis $\omega^{i} = (0, 0, \omega)$. We decompose $\rho$ in terms of Legendre polynomials as 
\begin{align}
  \rho = \sum_{\ell=0}^{\infty} \sqrt{\frac{2 \ell + 1}{4 \pi}} \rho_{\ell} P_{\ell}( \cos\theta ) ,
\end{align}
and retain up to the $\ell = 2$ mode of $\rho$, such that 
\begin{align}
  \rho = \frac{\rho_0}{\sqrt{4 \pi}} + \sqrt{\frac{5}{4 \pi}} \rho_2 P_2( \cos\theta ) , 
\end{align}
with $\rho_0$ and $\rho_2$ constant. Hereafter, we consider axisymmetric deformations of a sphere with constant density, such as a rotating ellipsoid, for which reflection symmetry with respect to the equatorial plane still exists. This reflection symmetry forces $\rho_{\ell} = 0$ for odd $\ell$.
The quantity $\rho_{0}$ represents the spherically symmetric component of the density distribution, while $\rho_{2}$ represents a quadrupolar deformation, for example induced by a second body. 

With this density profile at hand, we can then solve the Poisson equations for the PN potentials in Eq.~\eqref{eq:PN-Pot}. Doing so, $U$ and $V_k$ become~\cite{poisson2014gravity}
\begin{align}
  \label{Eq:UQ_1PN}
  & U =
    \begin{cases}
      \dfrac{M}{r} - J_2 \left( \dfrac{M}{r} \right) \left( \dfrac{R}{r} \right)^2 P_2( \cos\theta ) ,  & ( r > R ), \\
      \dfrac32 \dfrac{M}{R} \left( 1 - \dfrac13 \dfrac{r^2}{R^2} \right) - J_2 \dfrac{M}{R} \left( \dfrac{r}{R} \right)^2 \left( 1 - 5 \ln \dfrac{r}{R} \right) P_2( \cos \theta )  , & ( r \leq R ) , 
    \end{cases}
  \\
  \label{Eq:VQ_1PN}
  & V_k =
    \begin{cases}
      \vspace{0.2cm}
      \dfrac15 M R^2 \tilde{\epsilon}^0_{~ i j k} \omega^i \dfrac{x^j}{r^3} & ( r > R ) , \\
       \dfrac{\sqrt{\pi}}{3} \rho_0 \left( R^2 - \dfrac35 r^2 \right) \tilde{\epsilon}^0_{~ i j k} \omega^i x^j & ( r \leq R ), 
    \end{cases}
\end{align}
where total mass $M$ is defined by 
\begin{align}
  M \equiv \int \rho d^3 x 
  = \frac{\sqrt{4 \pi}}{3} \rho_0 R^3 ,
\end{align}
and we use the dimensionless quadrupole moment $J_2$ instead of $\rho_2$, where the former is defined by 
\begin{align}
  J_2 = - \frac15 \sqrt{\frac{4 \pi}{5}} \frac{R^3}{M} \rho_2 .
\end{align}

The cautious reader will notice that we have not included the $J_{2}$ corrections to the gravitomagnetic potential $V_{k}$ above. These corrections are generated by $\rho_2$ when solving Eq.~\eqref{eq:PN-Pot}. To be mathematically rigorous, such corrections would have to be included when computing the Pontryagin density because they enter at the same order as the $J_{2}$ corrections to the potential $U$. This can be seen from Eq.~\eqref{RR-PN}, using the $\ell=0$ part of $U$ and the $\ell=2$ part of $V_k$, which when included in Eq.~\eqref{gramag0} lead to effects of the same order-of-magnitude as those computed below. In the next section, however, we show that the $J_2$ correction is negligible in the analysis of the GPB experiment, and thus, henceforth we do not include the $J_2$ correction to $V_k$.

With the PN potentials at hand, we can now compute the evolution equation for the scalar field. Substituting these expressions into Eq.~\eqref{RR-PN}, the source term becomes 
\begin{align}
  \label{RRQPN}
  {^\ast\!} R R = 
  \begin{cases}
    \vspace{0.2cm}
    288 \dfrac{M^3 \chi}{r^7} \left[ 1 - \dfrac72 \dfrac{R^2}{r^2} J_2 \left( 1 + \dfrac57 \cos 2 \theta \right)  \right] \cos \theta + \mathcal{O} ( \varepsilon'^{\, 6}, \alpha' ) & ( r > R ) , \\
    1032 \dfrac{M^3 \chi }{R^7} \dfrac{r}{R} J_2 \left( 1 + \dfrac{60}{43} \ln \dfrac{r}{R} - \dfrac{15}{43} \cos 2 \theta \right) \cos \theta + \mathcal{O} ( \varepsilon'^{\, 6}, \alpha' ) & ( r \leq R ) , 
  \end{cases}
\end{align}
and the evolution equation [Eq.~\eqref{eq;scalar}] becomes 
\begin{align}
\label{scalarextintQ}
  \nabla^2 \vartheta =
  \begin{cases}
    \vspace{0.2cm}
    - 72 \sqrt{\dfrac{\zeta \kappa_g}{\beta}} \dfrac{M^2 R^3}{r^7} \chi \left[ 1 - \dfrac72 \dfrac{R^2}{r^2} J_2 \left( 1 + \dfrac57 \cos 2 \theta \right) \right] \cos \theta + \mathcal{O} ( \varepsilon'^{\, 6}, \alpha'^{\, 2} )  & ( r > R ), \\
    -258 \sqrt{\dfrac{\zeta \kappa_g}{\beta}} \dfrac{M^2 r}{R^5} \chi J_2 \left(1 + \dfrac{60}{43} \ln \dfrac{r}{R} - \dfrac{15}{43} \cos 2\theta \right) \cos\theta  +\mathcal{O} ( \varepsilon'^{\, 6}, \alpha'^{\, 2} ) & ( r \leq R )  . 
  \end{cases}
\end{align}
%%%%%%%%%%%%%%%%%%%%%%%%%%%%%%%%%%%%%%%%%%%%%%%%%%%%%
The integration constants can be determined by requiring finiteness of $\vartheta$ at $r=0$ and $r \to \infty$, as well as continuity and differentiability of the scalar field at the surface of the object. 
Imposing the above conditions, we obtain an axisymmetric scalar field for a weakly-gravitating and slowly-rotating object that is quadrupolarly deformed in the small-coupling and the PN approximations:
\begin{align}
  \label{Eq:thetaQ_ext}
  \vartheta_{\rm ext} =
  & 8 \sqrt{\dfrac{\zeta \kappa_g}{\beta}} \left( \dfrac{M}{r} \right)^3 \left( \frac{z}{M} \right) \chi \left[ 1 - \frac12 \frac{R^3}{r^3} - J_2 \left\{  1 - \frac{153}{98} \frac{R^2}{r^2} + \frac{255}{98} \frac{R^2}{r^2} \frac{z^2}{r^2} \left( 1 - \frac{49}{85} \frac{R^3}{r^3} \right) \right\} \right] \nonumber\\
  &+ \mathcal{O} ( \varepsilon'^{\, 6}, \alpha'^{\,2} ) ,  \\
  \label{Eq:thetaQ_int}
  \vartheta_{\rm int} =
  & 4 \sqrt{\dfrac{\zeta \kappa_g}{\beta}} \left( \dfrac{M}{R} \right)^3 \left( \frac{z}{M} \right) \chi \left[ 1 - \frac12 J_2 \left\{ 1 - \frac{159}{49} \frac{r^2}{R^2} \left( 1 - \frac{420}{53} \ln \frac{r}{R} \right) + \frac{216}{49} \frac{z^2}{R^2} \left( 1 - \frac{35}{12} \ln \frac{r}{R} \right) \right\} \right] \nonumber\\
  & + \mathcal{O} ( \varepsilon'^{\, 6}, \alpha'^{\,2} ) .
\end{align}
Clearly, this solution reduces to that in Eqs.~\eqref{Eq:theta_ext} and~\eqref{Eq:theta_int} in the $J_{2} \to 0$ limit, upon conversion of $\zeta$ to $\alpha$ via Eq.~\eqref{eq:zeta-def}. Notice also that in the exterior the $J_{2}$ contribution introduces new terms that fall off differently than in the spherically symmetric case. All of these contributions, however, are suppressed by $J_{2}$, which for most quadrupolar deformations is a small number. For example, the measured value of the non dimensional $J_2$ of the Earth is of
${\cal{O}}(10^{-3})$~\cite{murray2000solar}. We will see below that the slower decay rate of the $J_{2}$ part of the scalar field does not compensate for the smallness of $J_{2}$ because the experiments we are interested in here are conducted very close to the surface of Earth, for $r \sim R$. 

%----------------------------------------------------------------------------------------------------
\subsection{PN Solution for the dCS Metric Deformation: Gravitomagnetic Sector}

Let us now use the scalar field obtained in the previous section to source a metric deformation. The gravitomagnetic sector of the field equations was already presented in Eq.~\eqref{gramag0}. Substituting Eqs.~\eqref{1PN},~\eqref{Eq:UQ_1PN},~\eqref{Eq:thetaQ_ext}, and~\eqref{Eq:thetaQ_int} into Eq.~\eqref{gramag0}, we obtain
\begin{align}
  \label{eq;HQ-first}
  \nabla^2 H =
  \begin{cases}
    \vspace{2mm}
    144 \, \zeta \, \chi \, \dfrac{M^2 R^3}{r^7} \left[ 1 - \dfrac{R^3}{r^3} - J_2 \left\{ 1 - \dfrac{83}{98} \dfrac{R^2}{r^2} \left( 1 - \dfrac{931}{166} \dfrac{R^3}{r^3} \right) \right. \right. \nonumber \\
  \hspace{25mm}
  \left. \left. + \dfrac{515}{98} \dfrac{R^2}{r^2} \left(1 + \dfrac{49}{206} \dfrac{R^3}{r^3} \right) \cos 2 \theta \right\} \right] \sin \theta \, e^{- i \phi} + \mathcal{O} ( \varepsilon'^{\, 8}, \alpha'^{\, 3} , ( J_2 )^2 )  & ( r > R ) , \\
  600 \, \zeta \, \chi \dfrac{M^2}{r R^3} J_2 \, \cos^2 \theta \, \sin \theta \, e^{- i \phi} + \mathcal{O} ( \varepsilon'^{\, 8}, \alpha'^{\, 3}, ( J_2 )^2 )  & ( r \leq R ) ,
 \end{cases}
\end{align}
where, as before, we have defined $H \equiv \delta h_{0 y} + i \delta h_{0 x}$ with $i \equiv \sqrt{-1}$.

The gravitomagnetic field equations can now be solved in the same way as in the case of a spherically symmetric density profile. First, the z-component the metric deformation, $\delta h_{0z}$, must satisfy $\nabla^2\delta h_{0z}=0$, and by imposing regularity at the center of the object and asymptotic flatness at spatial infinity, we must require that $\delta h_{0z}=0$. 
%%%%%%%%%%%%%%%%%%%%%%%%%%%%%%%%%%%%%%%%%%%%%%%%%%%%%%%%%%%%%%%%%%%%
The integration constants can be determined by asymptotic flatness, regularity at the center and the matching condition at the surface. This time, however, the matching conditions are not trivial, but instead 
\begin{align}
  & \lim_{\lambda \to 0^{+}} \left[ H_{\rm ext} ( R + \lambda ) - H_{\rm int} ( R - \lambda ) \right] = 0 , \\
  & \lim_{\lambda \to 0^{+}} \left[ \left. \frac{d H_{\rm ext}}{d r} \right|_{r = R + \lambda} - \left. \frac{d H_{\rm int}}{d r} \right|_{r = R - \lambda} \right] = -24 \zeta \chi \frac{M^2}{R^3} \left[ 1 + \frac{1555}{294} J_2 \left( 1 - \frac{5619}{1555} \cos^2 \theta \right) \right] \sin\theta e^{-i\phi}
\end{align}
at leading PN order. 
Imposing these boundary conditions, we finally obtain
\begin{align}
  H_{\rm ext}
  &= 8 \, \zeta \, \chi \left( \frac{M}{r} \right)^2 \left( \frac{R}{r} \right)^2 \left[ \frac{R}{r} \left( 1 - \frac13 \frac{R^3}{r^3} \right) + \frac{87}{1715} J_2 \left\{ 1 - \frac{1715}{87} \frac{R}{r} \left( 1 - \frac{150}{49} \frac{R^2}{r^2} + \frac{201}{286} \frac{R^5}{r^5} \right) \right. \right. \nonumber \\
  &~~~ \left. \left. - 5 \left( 1 + \frac{721}{29} \frac{R^3}{r^3} + \frac{1715}{754} \frac{R^6}{r^6} \right) \cos^2 \theta \right\} \right] \sin \theta \, e^{- i \phi} + \mathcal{O} ( \varepsilon'^{\, 8}, \alpha'^{\, 3}, ( J_2 )^2 ) , \\
  H_{\rm int}
  &= \frac{16}{3} \, \zeta \, \chi \left( \frac{M}{R} \right)^2 \left( \frac{r}{R} \right) \left[ 1 +  \frac{489}{220} J_2 \left\{ 1 - \frac{35640}{726817} \frac{r^2}{R^2} + \frac{550}{163} \ln \left( \frac{r}{R} \right) \right. \right. \nonumber \\
  &~~~ \left. \left. - \frac{825}{163} \left( 1 - \frac{216}{4459} \frac{r^2}{R^2} \right) \cos^2 \theta \right\} \right] \sin \theta \, e^{- i \phi} + \mathcal{O} ( \varepsilon'^{\, 8}, \alpha'^{\, 3}, ( J_2 )^2 ) , 
\end{align}
and using the definition of $H$ we then find
\begin{align}
  \label{Eq:dhextx}
  \delta h^{\rm ext}_{0 x}
  &= - 8 \, \zeta \, \chi \left( \frac{M}{r} \right)^2 \left( \frac{R}{r} \right)^2 \left[ \frac{R}{r} \left( 1 - \frac13 \frac{R^3}{r^3} \right) + \frac{87}{1715} J_2 \left\{ 1 - \frac{1715}{87} \frac{R}{r} \left( 1 - \frac{150}{49} \frac{R^2}{r^2} + \frac{201}{286} \frac{R^5}{r^5} \right) \right. \right. \nonumber \\
  &~~~ \left. \left. - 5 \left( 1 + \frac{721}{29} \frac{R^3}{r^3} + \frac{1715}{754} \frac{R^6}{r^6} \right) \cos^2 \theta \right\} \right] \sin \theta \, \sin \phi + \mathcal{O} ( \varepsilon'^{\, 8}, \alpha'^{\, 3}, ( J_2 )^2 ) , \\
  \label{Eq:dhexty}
  \delta h^{\rm ext}_{0 y}
  &= 8 \, \zeta \, \chi \left( \frac{M}{r} \right)^2 \left( \frac{R}{r} \right)^2 \left[ \frac{R}{r} \left( 1 - \frac13 \frac{R^3}{r^3} \right) + \frac{87}{1715} J_2 \left\{ 1 - \frac{1715}{87} \frac{R}{r} \left( 1 - \frac{150}{49} \frac{R^2}{r^2} + \frac{201}{286} \frac{R^5}{r^5} \right) \right. \right. \nonumber \\
  &~~~ \left. \left. - 5 \left( 1 + \frac{721}{29} \frac{R^3}{r^3} + \frac{1715}{754} \frac{R^6}{r^6} \right) \cos^2 \theta \right\} \right] \sin \theta \, \cos \phi + \mathcal{O} ( \varepsilon'^{\, 8}, \alpha'^{\, 3}, ( J_2 )^2 ) , \\
  \delta h^{\rm int}_{0 x}
  &= - \frac{16}{3} \, \zeta \, \chi \left( \frac{M}{R} \right)^2 \left( \frac{r}{R} \right) \left[ 1 +  \frac{489}{220} J_2 \left\{ 1 - \frac{35640}{726817} \frac{r^2}{R^2} + \frac{550}{163} \ln \left( \frac{r}{R} \right) \right. \right. \nonumber \\
  &~~~ \left. \left. - \frac{825}{163} \left( 1 - \frac{216}{4459} \frac{r^2}{R^2} \right) \cos^2 \theta \right\} \right] \sin \theta \, \sin \phi + \mathcal{O} ( \varepsilon'^{\, 8}, \alpha'^{\, 3}, ( J_2 )^2 ) , \\
  \delta h^{\rm int}_{0 y} 
  &= \frac{16}{3} \, \zeta \, \chi \left( \frac{M}{R} \right)^2 \left( \frac{r}{R} \right) \left[ 1 +  \frac{489}{220} J_2 \left\{ 1 - \frac{35640}{726817} \frac{r^2}{R^2} + \frac{550}{163} \ln \left( \frac{r}{R} \right) \right. \right. \nonumber \\
  &~~~ \left. \left. - \frac{825}{163} \left( 1 - \frac{216}{4459} \frac{r^2}{R^2} \right) \cos^2 \theta \right\} \right] \sin \theta \, \cos \phi + \mathcal{O} ( \varepsilon'^{\, 8}, \alpha'^{\, 3}, ( J_2 )^2 ) , 
  \label{deltahQ}
\end{align}
As in the case of the scalar field, note that these solutions agree exactly with the spherically symmetric ones of the previous section in the $J_{2} \to 0$ limit. Observe also that the $J_{2}$ deformation introduces corrections that are parametrically small in $J_{2}$, but also introduce different fall-off rates in the exterior of the star. 

\begin{figure}[htbp]
  \begin{center}
    \includegraphics[width=120mm]{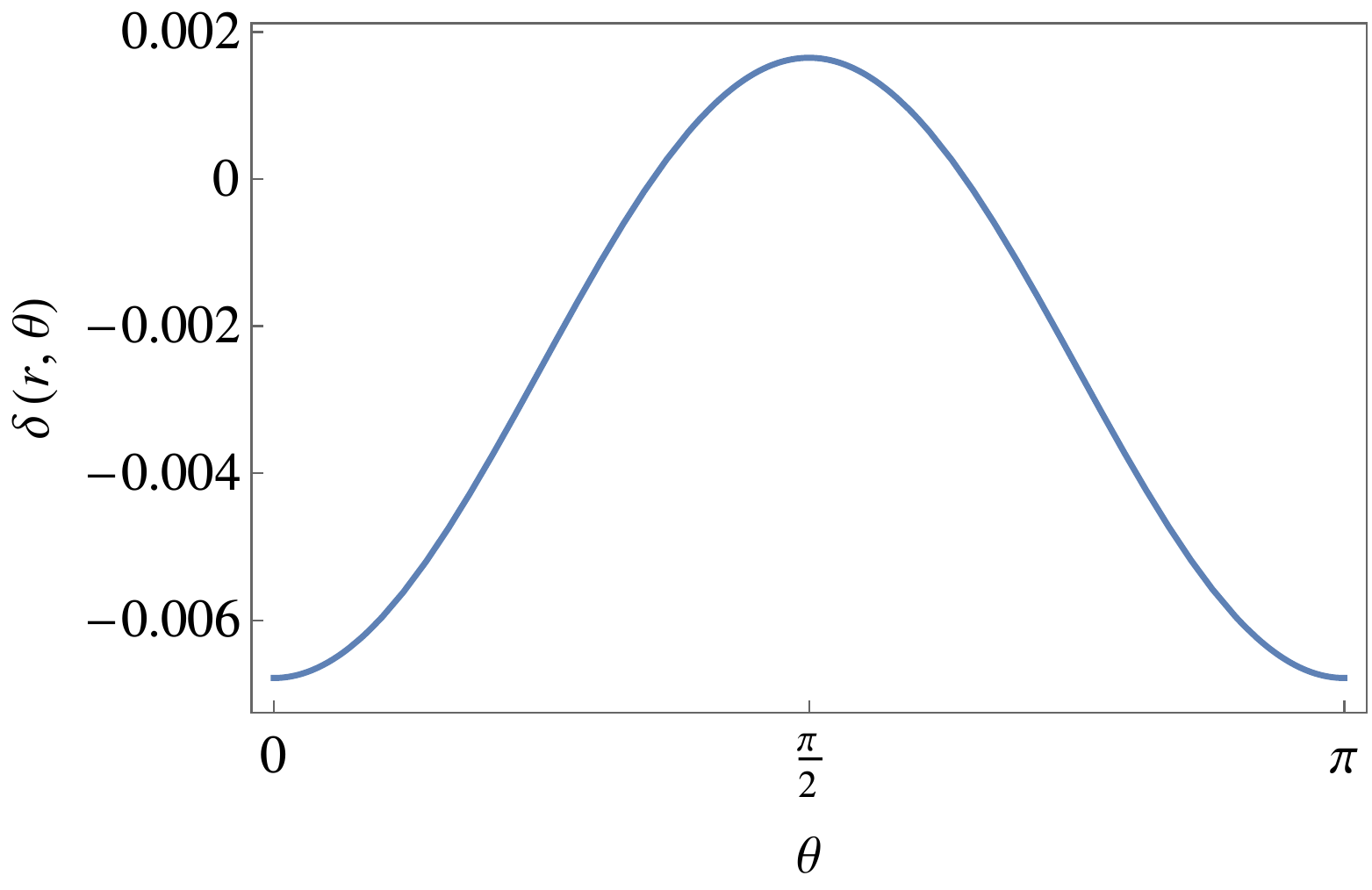}
  \end{center}
  \label{Fig:1}
  \caption{Deviation $\delta$ due to Earth's $J_2$ for the GPB gyroscopes as a function of latitude (polar) angle $\theta$. Observe that $\delta$ is very small, which implies that the $J_{2}$ corrections are three orders of magnitude smaller than the spherically symmetric terms.}
\end{figure}

Let us now study the magnitude of the $J_2$ corrections derived above. Consider the deviation from the spherically symmetric case,
\begin{align}
\delta ( r, \theta ) \equiv \frac{\delta h^{\rm ext}_{0 x} - \left( \delta h^{\rm ext}_{0 x} \right)_{J_2 \to 0}}{\delta h^{\rm ext}_{0 x}} = \frac{\delta h^{\rm ext}_{0 y} - \left( \delta h^{\rm ext}_{0 y} \right)_{J_2 \to 0}}{\delta h^{\rm ext}_{0 y}}.
\end{align}
Note that the deviation $\delta$ is independent of $\phi$ since Eqs.~\eqref{Eq:dhextx} and~\eqref{Eq:dhexty} depend on $\phi$ through the same overall factors, i.e.~$\sin \phi$ and $\cos \phi$.
Let us plot this quantity for the case of a satellite orbiting around Earth in Fig.~\ref{Fig:1}.
We have here used that Earth's $J_2 \sim 1.083 \times 10^{-3}$, that the GPB gyroscope is at a radius of $r \sim R + 650 \; {\rm{km}}$, and we have set $\phi = 0$.
Observe that regardless of the polar angle, the $J_{2}$ corrections are smaller than the spherical term by three orders of magnitude for GPB. Therefore, the $J_2$ corrections can be safely neglected in the GPB experiment. 

%%%%%%%%%%%%%%%%%%%%%%%%%%%%%%%%%%%%%%%%%%%%%%%%%
\section{Gyroscope Precession in the Gravity Probe B Experiment}
\label{Sec:GPB}

In this section, we discuss the spin precession of a gyroscope loaded on a satellite which orbits a weakly-gravitating and slowly-rotating object, namely the GPB experiment.
To leading order in the PN approximation, the precession of a spin three-vector ${\bm S}$ is expressed as~\cite{will1993theory} 
\begin{align}
  \label{spin}
  \frac{d {\bm S}}{d t} = {\bm \Omega} \times {\bm S} ,
\end{align}
where $\bm \Omega$ is the precession rate, which again to leading PN order can be obtained from the flat-space curl of the gravitomagnetic sector of the metric. The dominant effect of the dCS correction on the precession rate is thus produced by the gravitomagnetic components $\delta h_{0 i}$ via
\begin{align}
  \label{Eq:delta_Omega}
  \delta \bm{\Omega}
  &\equiv - \frac12 \nabla \times \delta \bm{g} = \frac{8 R^3}{r_{\rm gyro}^6} \zeta \left[ 2 \left( 1 - \frac{7}{12} \frac{R^3}{r_{\rm gyro}^3} \right) \bm{J} - 3 \left( 1 - \frac12 \frac{R^3}{r_{\rm gyro}^3} \right) \left(\bm{J} \cdot \bm{n}_{\rm gyro} \right) \bm{n}_{\rm gyro} \right] ,
\end{align}
where $\delta \bm{g} = \delta h_{0 i} \bm{e}_i$ with $\bm{e}_i$ a unit basis vector, the $\times$ operator is the outer product in flat space, $\bm{J} = (0, 0, J)$ is the spin angular momentum vector of the gravitating object, and $r_{\rm gyro}$ and $\bm{n}_{\rm gyro}$ are the distance of the gyroscope from the center of the object and the unit vector from the object to the gyroscope, respectively. This expression neglects the corrections to the precession rate induced by a quadrupolar deformation of the gravitating object, since as we saw in the previous section, these are negligibly small.

Since the characteristic time-scale of spin precession is much larger than the satellite's orbital period $T_{\rm gyro}$, let us average Eq.~\eqref{spin} over the latter. Introducing the normalized spin vector $\hat{\bm{S}} \equiv \bm{S}/|\bm{S}|$, the precession rate of the spin vector can be written as
\begin{align}
  \left\langle \bm{P}_r \right\rangle_{T_{\rm gyro}} \equiv 
  \frac{1}{T_{\rm gyro}} \int_0^{T_{\rm gyro}} d t ( \bm{\Omega} \times \hat{\bm{S}} ) 
  = 
  \frac{1}{T_{\rm gyro}} 
  \left( \int_0^{T_{\rm gyro}} \bm{\Omega} d t \right) \times \hat{\bm{S}} ,
  \label{Eq:dS_average}
\end{align}
Substituting Eq.~\eqref{Eq:delta_Omega} into Eq.~\eqref{Eq:dS_average}, we obtain the dCS correction to the averaged rate of spin precession:
\begin{align}
  \left\langle \delta \bm{P}_r \right\rangle_{T_{\rm gyro}} = \frac{4 R^3}{r_{\rm gyro}^6} \zeta \left[ \left( 1 - \frac56 \frac{R^3}{r_{\rm gyro}^3} \right) \bm{J} + 3 \left( 1 - \frac12 \frac{R^3}{r_{\rm gyro}^3} \right) \left(\bm{J} \cdot \bm{h}_{\rm gyro} \right) \bm{h}_{\rm gyro} \right] \times \hat{\bm{S}} ,
  \label{Eq:Pr_dCS}
\end{align}
where $\bm{h}_{\rm gyro}$ is the unit vector of the satellite's orbital angular momentum. Equation~\eqref{Eq:Pr_dCS} shows that, even in the case of a single star, the leading-order dCS correction appears in the same direction as the Lense-Thirring effect in GR~\cite{will1993theory}, i.e.~in the so-called West-East (WE) direction (see \ref{App:GyroCoord}). This is in stark contrast to the prediction of non-dynamical Chern-Simons gravity~\cite{Alexander:2007vt}.

\begin{figure}[htbp]
  \begin{center}
    \includegraphics[width=120mm]{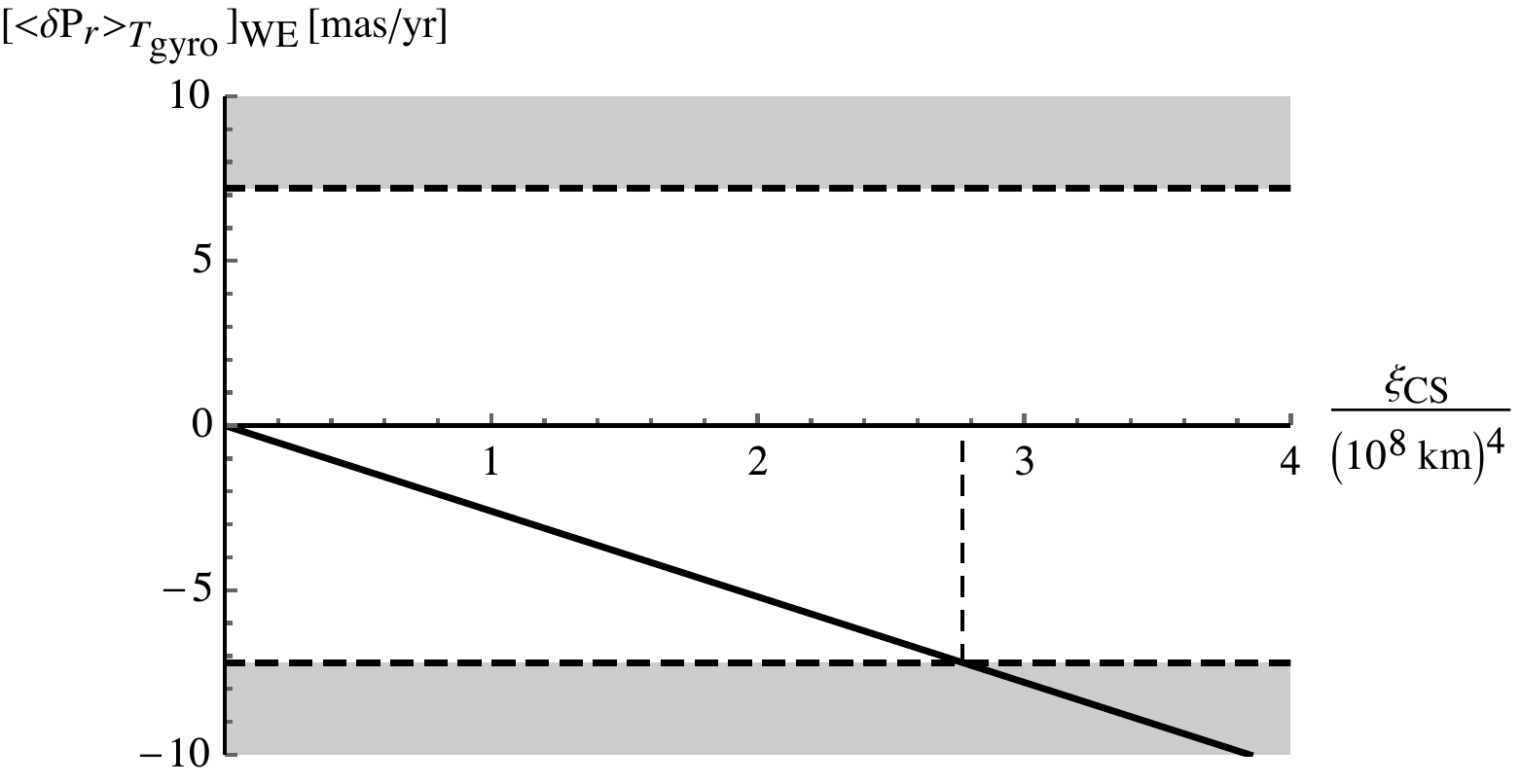}
  \end{center}
  \label{Fig:2}
  \caption{DCS correction to the spin precession rate of GPB gyroscopes in the WE direction as a function of the $\xi_{CS}$ parameter. Given that GPB verified the predictions of GR to an accuracy of about $\pm 7.2$ mas/yr~\cite{Everitt:2011hp} (horizontal dashed curves), dCS gravity is constrained to $\xi_{CS} \lesssim 2.8 \times (10^{8} \; {\rm{km}})^{4}$. The shaded regions in this figure are inconsistent with the GPB experiment. 
  }
\end{figure}

Let us now focus on the GPB experiment. For the parameters associated with that experiment, the dCS correction to the spin precession rate in the WE direction becomes
\begin{align}
  \left( \left\langle \delta \bm{P}_r \right\rangle_{T_{\rm gyro}} \right)_{\rm WE} \simeq - 2.6 \left( \frac{\xi_{\rm CS}}{\left( 10^8 \, {\rm km} \right)^4} \right) \,\, {\rm [mas/year]} .
  \label{Eq:Pr_dCS_WE}
\end{align}
This expression is plotted in Fig.~\ref{Fig:2} as a function of the $\xi_{CS}$ parameter for the GPB experiment, where the horizontal, dashed curves correspond to the accuracy to which GPB verified the predictions of GR. The region to the right of the intersection of the solid curve and the dashed curve is then ruled out by the GPB experiment. Numerically, this corresponds to a bound of 
\[
  \xi_{\rm CS}^{1/4} \lesssim 1.3 \times 10^8 \, {\rm km} ,
\]
which is consistent with the order of magnitude estimate derived in~\cite{AliHaimoud:2011fw} by studying the dCS metric perturbation. This consistency is reasonable because the gravitomagnetic components of the dCS corrections in Eqs.~\eqref{Eq:delta_h0x_ext} and~\eqref{Eq:delta_h0y_ext} correspond to a special case of the solution discussed in~\cite{AliHaimoud:2011fw}, and it is these components that induce a modification to the GPB precession rate. One should mention, however, that the formalism of~\cite{AliHaimoud:2011fw}, the so-called Hartle-Thorne formulation, is only capable of handling a single body, while our formulation is more general, being applicable to other scenarios, such as binary systems, $N$-body systems, and isolated bodies that are not spherically symmetric. 

Let us conclude with an order of magnitude estimate of the contribution of the temporal-temporal component of the dCS metric deformation to the spin-precession correction calculated above. The temporal-temporal component of the dCS metric deformation scales as
\begin{align}
  \delta h_{0 0} \sim \frac{M J^2}{R^6 r^3} \xi_{\rm CS},
\end{align}
and thus the precession rate due to this term can be estimated to be 
\begin{align}
  \delta \Omega \sim v_{\rm gyro} \delta h_{0 0} \sim 10^{-11} \left( \frac{\xi_{\rm CS}}{\left( 10^8 \, {\rm km} \right)^4} \right) \, {\rm [mas/year]} .
\end{align}
One can clearly see that this effect is much smaller than the contribution from the gravitomagnetic components [Eq.~\eqref{Eq:Pr_dCS_WE}]. One can also easily show that the purely spatial components also lead to modifications to the precession rate that are of the same order as those coming from the temporal-temporal deformation. This thus proves that one can ignore the $\delta h_{00}$ and $\delta h_{i j}$ contributions to spin precession in dCS gravity.

%%%%%%%%%%%%%%%%%%%%%%%%%%%%%%%%%%%%%%%%%%%%%%%%%
\section{Conclusion}
\label{Sec:Conclution}

We have calculated the interior and exterior profiles of the scalar field and metric deformation in dCS gravity for a slowly-rotating and weakly-gravitating object of uniform mass density and for an object with a quadrupolar deformation using the PN and the small-coupling approximations. We have found that the asymptotic, peeling behavior of the exterior solutions is consistent with that found in previous works concerning black holes~\cite{Yunes:2009hc, Yagi:2012ya} and neutron stars~\cite{Yagi:2013mbt}, as well as non-relativistic objects~\cite{AliHaimoud:2011fw}. We then used these solutions to evaluate the dCS correction to spin precession, as would be applicable for example to the GPB experiment. The verification of Lense-Thirring precession by the latter then allows for a constraint on the dCS coupling parameter, which are comparable to bounds on the non-dynamical theory~\cite{AliHaimoud:2011fw}. We have found that the dCS correction to spin-precession does not vanish, and is in fact, in the same direction as the Lense-Thirring effect in GR. 

Although we focus on isolated objects, our approach is easily applicable to more complex scenarios, like binary systems and $N$-body systems of weakly-gravitating bodies. Future work could concentrate on extending this analysis to binary systems or to $N$-bodies. Provided these bodies are weakly-gravitating and slowly-spinning, then the formalism applied here should be easily extendible. Ultimately, however, this analysis makes it painfully clear that meaningful constraints on dCS gravity can only come from observations that sample the extreme gravity regime, where the gravitational interaction is non-linear, dynamical and strong, which is also where dCS corrections will be largest. Any Solar System constraint is bound to be restricted by the feebleness of the gravitational interaction in the Solar neighborhood.    

%%%%%%%%%%%%%%%%%%%%%%%%%%%%%%%%%%%%%%%%%%%%%%%%%
\section*{Acknowledgments}
The authors are grateful to Takahiro Tanaka for incisive comments and stimulating discussions.
Y.N., D.K., K.Y., and H.A. wish to thank Riko Yamada and Yuuiti Sendouda for useful discussions.
This work was supported in part by Japan Society for the Promotion of Science (JSPS) Grant-in-Aid for Scientific Research, No. 26400262 (H.A.), No. 17K05431 (H.A.), by the Ministry of Education, Culture, Sports, Science, and Technology, No. 15H00772 (H.A.) and No. 17H06359 (H.A.), by the JSPS Grant-in-Aid for JSPS Fellows, No. 15J01732 (K.Y.).
N. Y. acknowledge support from NSF EAPSI Award No. 1614203, NSF CAREER grant PHY-1250636, and NASA grants NNX16AB98G and 80NSSC17M0041.

%%%%%%%%%%%%%%%%%%%%%%%%%%%%%%%%%%%%%%%%%%%%%%%%%
\appendix
\section{Linearized Ricci tensor}
\label{App:LR-tensor}

\if0%%%%%%%%%%%%%%%%%%%%%%%%%%%%%%%%%%%%%%%%%%%%%%%%%%%%%%%%
The metric perturbation $h_{\mu \nu}$ included C-tensor is evaluated by 1PN metric. Therefore we use the PN gauge for is expanded C-tensor.
By using the PN gauge, the C-tensor reduces to   
\begin{align}
\label{Ccomp}
C_{0 i}=&
\frac14\bigg[2\tilde{\epsilon}^{0ljk}\partial_m\{\vartheta_{,l}h_{k[m,i]j}\}-\tilde{\epsilon}^{0 i j k}\partial_l\{\vartheta_{,j}h_{00,kl}\}\bigg]
+\mathcal{O}(\varepsilon'^{\, 8}, \alpha'^{\, 2}) , \nonumber \\
C_{0 0}=&-\frac12\vartheta_{,i}\tilde{\epsilon}^{0 i j k}
\nabla^2 h_{0k,j}
+\frac12\vartheta_{,mi}\tilde{\epsilon}^{0 i j k}(h{}^{m}_{k,j0}-h_{0k,j}{}^{m}) +\mathcal{O}(\varepsilon'^{\, 9},\alpha'^{\, 2}) , 
\nonumber \\
C_{i j}=&\frac12\vartheta_{,m}\tilde{\epsilon}^{0mn}{}_{(i}
(\nabla^2 h_{j)n,0}-h_{ln,0|j)}{}^{l})
-\frac12\vartheta_{,m}\tilde{\epsilon}^{0mn}{}_{(i}
(\nabla^2 h_{j)0,n}-h_{00,n|j)}{}^{0}-h_{l0,n|j)}{}^l)  \nonumber \\
&-\vartheta_{,lm}\tilde{\epsilon}^{0mn}{}_{(i}(h{}^l{}_{n,0|j)}-h_{j)n,0}{}^l)+\mathcal{O}(\varepsilon'^{\, 9},\alpha'^{\, 2}) .
\end{align}
Note that the gravitomagnetic components of the dCS metric deformation are the largest. 
\fi%%%%%%%%%%%%%%%%%%%%%%%%%%%%%%%%%%%%%%%%%%%%%%%%%%%%%%%%

The Ricci tensor, linearized around a Minkowski background in GR, is
\begin{align}
\delta R_{\mu \nu} [h] = - \frac12 \left[ \Box_{\eta} h_{\mu \nu} - 2 h_{\alpha ( \mu, \nu )}^{~~~~~~ \alpha} + h_{, \mu \nu}\right] + \mathcal{O} ( h^2 ) , 
\end{align}
where $\Box_{\eta}$ is the d'Alembertian operator in flat spacetime. Since the time derivatives are higher order than spatial ones in the PN approximation, at leading PN order, and without imposing any gauge conditions on $h_{\mu \nu}$, the components of the linearized Ricci tensor in Cartesian coordinates are
\begin{align*}
\delta R_{0 0} =& - \frac12 ( \nabla^2 h_{0 0} - 2 h_{i 0 , 0}^{~~~\, i} + h^i_{~ i , 0 0} ) + \mathcal{O} ( h^2 ) , \\
\delta R_{0 i} =& - \frac12 ( \nabla^2 h_{0 i} - h^j_{0 , i j} - h^j_{i , 0 j} + h^j_{j , 0 i} ) + \mathcal{O} ( h^2 ) , \\
\delta R_{i j} =& - \frac12 ( \nabla^2 h_{i j} + h_{i j, 0}^{~~~\, 0} - h^0_{i , j 0} - h^k_{~ i , j k} - h^0_{~j , i 0} - h^k_{~ j, i k} + h^0_{~ 0, i j} + h^k_{~ k, i j} ) + \mathcal{O} ( h^2 ) .
\end{align*}
In order to obtain the linearized Ricci tensor evaluated on the dCS metric deformation, we can use the above expression with the replacement $h_{\mu \nu} \to \delta h_{\mu \nu}$. Imposing the standard PN gauge conditions on $\delta h_{\mu \nu}$, namely Eqs.~\eqref{Eq:ModPNgauge1} and~\eqref{Eq:ModPNgauge2}, one obtains Eqs.~\eqref{Eq:deltaR00}-\eqref{Eq:deltaRij}. There are other gauge choices that would lead to different expressions for the linearized Ricci tensor. 
\if0
\begin{align}
&R_{0 0}=-\frac12\nabla^2 h_{0 0}+\mathcal{O}(\varepsilon'^{\, 9},\alpha'^{\, 2}) , \nonumber \\
&R_{0 i}=-\frac12 \nabla^2 h_{0 i}+\mathcal{O}(\varepsilon'^{9}\alpha'^{\, 2}) , \nonumber \\
&R_{i j}=-\frac12\nabla^2 h_{i j}
-h_{0(i,j)0}+\mathcal{O}(\varepsilon'^{\, 9},\alpha'^{\, 2}) . \nonumber
\end{align}
\fi

%%%%%%%%%%%%%%%%%%%%%%%%%%%%%%%%%%%%%%%%%%%%%%%%%
\section{Details of the Derivations of the Weak Field Solution with Quadrupolar Deformation}
\label{App:FullCal}

In this appendix, we provide details of the derivation and expressions for the scalar field and gravitomagnetic sector of the metric, which are decomposed by Legendre polynomials $P_{\ell} (\cos\theta)$ and spherical harmonics $Y_{\ell}{}^{m} (\theta, \phi)$, respectively. We solve Eq.~\eqref{scalarextintQ} through a decomposition of $\vartheta(r, \theta)$ in terms of Legendre polynomials, as in Eq.~\eqref{eq:Leg-Dec-theta}. The source term in Eq.~\eqref{scalarextintQ} vanishes at leading PN order when $\ell \neq 1$ and $\ell \neq 3$, and thus, we focus on the $\ell = 1$ and $\ell = 3$ modes only, for which this equation becomes
\begin{align}
  \label{eq;SQ1-PN}
  \left[ \frac{1}{r^2} \frac{d}{d r} \left( r^2 \frac{\partial}{\partial r} \right) - \frac{2}{r^2} \right] \vartheta_1 =
  \begin{cases}
    \vspace{0.2cm}
   - 72 \sqrt{\dfrac{\zeta \kappa_g}{\beta}} \dfrac{M^2 R^3}{r^7} \chi \left( 1 - 4 \dfrac{R^2}{r^2} J_2 \right) + \mathcal{O}( \varepsilon'^{\, 6}, \alpha'^{\, 2} ) & ( r > R ) , \\
   - 240 \sqrt{\dfrac{\zeta \kappa_g}{\beta}} \dfrac{M^2 r}{R^5} \chi J_2 \left( 1 + \dfrac32 \ln \dfrac{r}{R}\right) + \mathcal{O}( \varepsilon'^{\, 6}, \alpha'^{\, 2} ) & ( r \leq R ) , 
  \end{cases}
\end{align} 
and
\begin{align}
   \label{eq;SQ3-PN}
   \left[ \frac{1}{r^2} \frac{d}{d r} \left( r^2 \frac{\partial}{\partial r} \right) - \frac{12}{r^2} \right] \vartheta_3 =
  \begin{cases}
    \vspace{0.2cm}
   144 \sqrt{\dfrac{\zeta \kappa_g}{\beta}} \dfrac{M^2 R^5}{r^9} \chi J_2 + \mathcal{O}( \varepsilon'^{\, 6}, \alpha'^{\, 2} ) & ( r > R ) , \\
  72 \sqrt{\dfrac{\zeta \kappa_g}{\beta}} \dfrac{M^2 r}{R^5} \chi J_2 + \mathcal{O}( \varepsilon'^{\, 6}, \alpha'^{\, 2} ) & ( r \leq R ) , 
  \end{cases}
\end{align}
The general solutions to these equations are straightforward and we find
\begin{align}
  \vartheta_1 = 
  \begin{cases}
    \vspace{0.2cm}
    \mathcal{A}_1 r + \dfrac{\mathcal{A}_2}{r^2} -4 \sqrt{\dfrac{\zeta \kappa_g}{\beta}} \dfrac{M^2 R^3}{r^5} \chi \left(1 - \dfrac{9 J_2}{5} \dfrac{R^2}{r^2} \right)  + \mathcal{O}( \varepsilon'^{\, 6}, \alpha'^{\, 2} ) & ( r > R ) , \\
    \mathcal{B}_1 r + \dfrac{\mathcal{B}_2}{r^2}+ \dfrac{6}{5} \sqrt{\dfrac{\zeta \kappa_g}{\beta}} \dfrac{M^2 r^3}{R^5} \chi J_2 \left(1 - 30 \ln \dfrac{r}{R} \right)+\mathcal{O}( \varepsilon'^{\, 6}, \alpha'^{\, 2} ) & ( r \leq R ) , 
  \end{cases}
\end{align}
for the $\ell = 1$ mode, and
\begin{align}
  \label{eq:SolThetaEll3}
  \vartheta_3 = 
  \begin{cases}
    \vspace{0.2cm}
    \mathcal{A}_3 r^3 + \dfrac{\mathcal{A}_4}{r^4}+ \dfrac{24}{5}\sqrt{\dfrac{\zeta \kappa_g}{\beta}} \dfrac{M^2 R^5}{r^7} \chi J_2 + \mathcal{O}( \varepsilon'^{\, 6}, \alpha'^{\, 2} ) & ( r > R ) , \\
    \mathcal{B}_3 r^3 + \dfrac{\mathcal{B}_4}{r^4} - \dfrac{72}{49} \sqrt{\dfrac{\zeta \kappa_g}{\beta}} \dfrac{M^2 r^3}{R^5} \chi J_2 \left(1- 7 \ln \dfrac{r}{R} \right)+\mathcal{O}( \varepsilon'^{\, 6}, \alpha'^{\, 2} ) & ( r \leq R ) , 
  \end{cases}
\end{align}
for the $\ell = 3$ mode, where $\mathcal{A}_i$ and $\mathcal{B}_i$ ($i = 1, 2, 3, 4$) are integration constants.

Just as in the spherically symmetric case, we determine these integration constants by requiring regularity and smoothness of the solution. By requiring that $\vartheta$ be finite as $r \to 0$ and $r \to \infty$, $\mathcal{A}_1$, $\mathcal{A}_3$, $\mathcal{B}_2$ and $\mathcal{B}_4$ must all vanish. The remaining four constants can be determined by requiring continuity and differentiability at the surface of the object, namely,
\begin{align}
  \lim_{\lambda \to 0^{+}} \left[ \vartheta_{\rm ext} ( R + \lambda ) - \vartheta_{\rm int} ( R - \lambda ) \right] = 0 ,  \\
  \lim_{\lambda \to 0^{+}} \left[ \left. \frac{\partial \vartheta_{\rm ext}}{\partial r} \right|_{r = R + \lambda} - \left. \frac{\partial \vartheta_{\rm int}}{\partial r} \right|_{r = R - \lambda} \right] = 0. 
\end{align}
One can verify that the latter condition holds by integrating Eq.~\eqref{eq;SQ1-PN} and Eq.~\eqref{eq;SQ3-PN}. Imposing the above conditions, we obtain 
\begin{align}
&\vartheta_1=
 \begin{cases}
    \vspace{0.2cm}
 8 \sqrt{\dfrac{\zeta \kappa_g}{\beta}} \bigg(\dfrac{M}{r} \bigg)^2 \chi \bigg[1- \dfrac{1}{2} \dfrac{R^3}{r^3} 
- J_2 \bigg(1 - \dfrac{9}{10} \dfrac{R^5}{r^5}  \bigg)  \bigg] + \mathcal{O}( \varepsilon'^{\, 6}, \alpha'^{\, 2} ) & ( r > R ) , \\
 4 \sqrt{\dfrac{\zeta \kappa_g}{\beta}} \bigg(\dfrac{M}{R} \bigg)^2 \dfrac{r}{R} \chi \bigg[
 1 + \dfrac{3}{10} \dfrac{r^2}{R^2} \bigg(1 - \dfrac{5}{3} \dfrac{R^2}{r^2} - 30 \ln \dfrac{r}{R}  \bigg)\bigg]
 +\mathcal{O}( \varepsilon'^{\, 6}, \alpha'^{\, 2} ) & ( r \leq R ) , \\
    \end{cases}   \\
&\vartheta_3=
 \begin{cases}
    \vspace{0.2cm}
  - \dfrac{408}{49} \sqrt{\dfrac{\zeta \kappa_g}{\beta}} \bigg( \dfrac{M}{r} \bigg)^2 \bigg( \dfrac{R}{r} \bigg)^2 \chi J_2 \bigg(1 - \dfrac{49}{85} \dfrac{R^3}{r^3} \bigg)+ \mathcal{O}( \varepsilon'^{\, 6}, \alpha'^{\, 2} ) & ( r > R ) , \\
 - \dfrac{864}{245} \sqrt{\dfrac{\zeta \kappa_g}{\beta}} \bigg(\dfrac{M}{R}\bigg)^2 \bigg( \dfrac{r}{R} \bigg)^3 \bigg[1 - \dfrac{35}{12} \ln \dfrac{r}{R} \bigg]+\mathcal{O}( \varepsilon'^{\, 6}, \alpha'^{\, 2} ) & ( r \leq R ) . 
\end{cases}
\end{align}

The gravitomagnetic sector of the field equation can be written as 
\begin{align}
  \label{eq;HQ-second}
  \nabla^2 H =
  \begin{cases}
    \displaystyle
     144 \zeta \chi \dfrac{M^2 R^3}{r^7} \bigg[1 - \dfrac{R^3}{r^3} - J_2 \bigg\{1 - \dfrac{83}{98} \dfrac{R^2}{r^2} + \dfrac{19}{4} \dfrac{R^5}{r^5} + \dfrac{515}{98} \dfrac{R^2}{r^2} \cos 2\theta + \dfrac{5}{4} \dfrac{R^5}{r^5} \cos 2\theta 
\\+ \dfrac{19 J_2}{4} \dfrac{R^2}{r^2}  \bigg(1 + \dfrac{5}{19} \cos 2 \theta  - \dfrac{21675}{7448} \dfrac{R^2}{r^2} - \dfrac{105}{19} \dfrac{R^5}{r^5} - \dfrac{2295}{1862} \dfrac{R^2}{r^2} \cos 2\theta - \dfrac{120}{19} \dfrac{R^5}{r^5} \cos 2\theta 
\\- \dfrac{255}{1064} \dfrac{R^2}{r^2} \cos 4\theta - \dfrac{15}{19} \dfrac{R^5}{r^5} \cos 4\theta   \bigg)\bigg\}  \bigg] \sin\theta e^{-i \phi} + \mathcal{O} (\varepsilon'^{\, 8}, \alpha'^{\, 3})  ( r > R ) , \\
    \displaystyle
    300 \zeta \chi \dfrac{M^2}{r R^3} J_2 \bigg\{1 + \cos 2\theta - \dfrac{J_2}{2}\bigg(1 + \dfrac{31931}{1225} \dfrac{r^2}{R^2} + \dfrac{36297}{490} \dfrac{r^2}{R^2} \ln\dfrac{r}{R} + \dfrac{144}{7} \dfrac{r^2}{R^2} \bigg(\ln \dfrac{r}{R} \bigg)^2 
    \\+\cos 2\theta +\dfrac{6261}{245} \dfrac{r^2}{R^2} \cos 2\theta + \dfrac{18}{35} \dfrac{r^2}{R^2} \cos 4\theta 
    \\+ \dfrac{360}{7} \dfrac{r^2}{R^2} \ln \dfrac{r}{R} \cos 2\theta - \dfrac{171}{14} \dfrac{r^2}{R^2} \ln\dfrac{r}{R} \cos 4\theta \bigg)   \bigg\} \sin\theta e^{-i\phi} + \mathcal{O} (\varepsilon'^{\, 8}, \alpha'^{\, 3})   ( r \leq R ) ,
 \end{cases}
\end{align}
up to ${\cal{O}}(J_2^2)$. Decomposing $H$ into  spherical harmonics $Y_{\ell}^m(\theta, \phi)$ as in Eq.~\eqref{eq:H-decomp}, we note that the only non-vanishing modes are $(\ell,m)=(1,-1)$, $(\ell,m)=(3,-1)$, and $(\ell,m)=(5,-1)$. These modes must satisfy 
\allowdisplaybreaks[4]
\begin{align}
  \label{eq;HQ1-1}
&  \frac{1}{r^2} \frac{d}{d r} \left( r^2 \frac{d H_1^{- 1}}{d r} \right) - 2 \frac{H_1^{- 1}}{r^2} =
  \begin{cases}
    \displaystyle
    96 \sqrt{6 \pi} \zeta \chi \dfrac{M^2 R^3}{r^7} \bigg[1 - \dfrac{R^3}{r^3} - J_2 \bigg\{1 -4 \dfrac{R^2}{r^2} +4 \dfrac{R^5}{r^5} 
    \\+4J_2 \dfrac{R^2}{r^2} \left(1 - \dfrac{255}{98} \dfrac{R^2}{r^2} - \dfrac{15}{7} \dfrac{R^5}{r^5}\right)   \bigg\} \bigg] + \mathcal{O} (\varepsilon'^{\, 8}, \alpha'^{\, 3}) & ( r > R ) , \\
    \displaystyle
    80 \sqrt{6 \pi} \zeta \chi \dfrac{M^2}{r R^3} J_2 \bigg\{1 - \dfrac{J_2}{2} \left(1 + \dfrac{943}{35} \dfrac{r^2}{R^2} + \dfrac{738}{7} \dfrac{r^2}{R^2} \ln  \dfrac{r}{R}
\right.
\\
\left.
  + \dfrac{360}{7} \dfrac{r^2}{R^2} \left(\ln \dfrac{r}{R}\right)^2  \right)  \bigg\} +\mathcal{O} (\varepsilon'^{\, 8}, \alpha'^{\, 3}) & ( r \leq R ) ,
  \end{cases}
\\
  \label{eq;HQ3-1}
&  \frac{1}{r^2} \frac{d}{d r} \left( r^2 \frac{d H_3^{- 1}}{d r} \right) - 12 \frac{H_3^{- 1}}{r^2} =
  \begin{cases}
    \displaystyle
     - \dfrac{39552}{49} \sqrt{\dfrac{3 \pi}{7}} \zeta \chi \dfrac{M^2 R^5}{r^9} J_2 \bigg\{1 + \dfrac{49}{206} \dfrac{R^3}{r^3}  
     \\ +\dfrac{49 J_2}{206} \left(1 - \dfrac{170}{49} \dfrac{R^2}{r^2} - 20 \dfrac{R^5}{r^5}\right)\bigg\}+ \mathcal{O} (\varepsilon'^{\, 8}, \alpha'^{\, 3}) & ( r > R ) , \\
    \displaystyle
    320 \sqrt{\dfrac{3 \pi}{7}} \zeta \chi \dfrac{M^2}{r R^3} J_2 \bigg\{1 - \dfrac{J_2}{2} \left(1 + \dfrac{6093}{245} \dfrac{r^2}{R^2}
      \right.
      \\
      \left.
        + \dfrac{474}{7} \dfrac{r^2}{R^2} \ln  \dfrac{r}{R}  \right)  \bigg\}  +\mathcal{O} (\varepsilon'^{\, 8}, \alpha'^{\, 3}) & ( r \leq R ) ,
  \end{cases}
\\
  \label{eq;HQ5-1}
&  \frac{1}{r^2} \frac{d}{d r} \left( r^2 \frac{d H_5^{- 1}}{d r} \right) - 30 \frac{H_5^{- 1}}{r^2} =
  \begin{cases}
    \displaystyle
     \dfrac{3264}{49} \sqrt{\dfrac{30 \pi}{11}} \zeta \chi \dfrac{M^2 R^7}{r^{11}} (J_2)^2 \left(1 + \dfrac{56}{17} \dfrac{R^3}{r^3}\right)+ \mathcal{O} (\varepsilon'^{\, 8}, \alpha'^{\, 3}) & ( r > R ) , \\
    \displaystyle
     - \dfrac{1536}{49} \sqrt{\dfrac{30 \pi}{11}} \zeta \chi \dfrac{M^2 r}{R^5}  (J_2)^2 \left(1 - \dfrac{95}{4} \ln  \dfrac{r}{R}  \right) +\mathcal{O} (\varepsilon'^{\, 8}, \alpha'^{\, 3}) & ( r \leq R ) ,
  \end{cases}
\end{align}
up to ${\cal{O}}(J_2^2)$. The general exterior and interior solutions are then
\begin{align}
&  H_1^{-1} = 
  \begin{cases}
  \displaystyle 
  \mathcal{C}_1 r + \dfrac{\mathcal{C}_2}{r^2} + 16 \sqrt{\dfrac{2 \pi}{3}} \zeta \chi \left(\dfrac{M}{r}\right)^2 \left(\dfrac{R}{r}\right)^3 \bigg[1 - \dfrac{1}{3} \dfrac{R^3}{r^3} 
  \\- J_2 \bigg\{1 - \dfrac{9}{5} \dfrac{R^2}{r^2} + \dfrac{9}{11} \dfrac{R^5}{r^5} + \dfrac{9 J_2}{5} \dfrac{R^2}{r^2} \left(1 - \dfrac{510}{343} \dfrac{R^2}{r^2} - \dfrac{60}{91} \dfrac{R^5}{r^5}  \right)  \bigg\}  \bigg] + \mathcal{O} (\varepsilon'^{\, 8}, \alpha'^{\, 3}) & ( r > R ) , \\
  \displaystyle 
   \mathcal{D}_1 r + \dfrac{\mathcal{D}_2}{r^2} - \dfrac{80}{3} \sqrt{\dfrac{2 \pi}{3}} \zeta \chi \left(\dfrac{M}{R}\right)^2 \left(\dfrac{r}{R}\right) J_2 \bigg\{1 -3 \ln  r 
   \\- \dfrac{J_2}{2} \left(1 + \dfrac{1062}{175} \dfrac{r^2}{R^2} - 3 \ln  r - \dfrac{1053}{35} \dfrac{r^2}{R^2} \ln  \dfrac{r}{R} - \dfrac{324}{7} \dfrac{r^2}{R^2} \left(\ln  \dfrac{r}{R} \right)^2 \right)  \bigg\} + \mathcal{O} (\varepsilon'^{\, 8}, \alpha'^{\, 3}) & ( r \leq R ) ,
  \end{cases}
\\
&  H_3^{-1} = 
  \begin{cases}
  \displaystyle 
   \mathcal{C}_3 r^3 + \dfrac{\mathcal{C}_4}{r^4} - \dfrac{6592}{245} \sqrt{\dfrac{3 \pi}{7}} \zeta \chi \left(\dfrac{M}{r}\right)^2  \left(\dfrac{R}{r}\right)^5 J_2 \bigg\{1 + \dfrac{245}{2678} \dfrac{R^3}{r^3} 
   \\+ \dfrac{49 J_2}{206} \left(1 - \dfrac{85}{49} \dfrac{R^2}{r^2} - 5 \dfrac{R^5}{r^5}  \right)  \bigg\}+ \mathcal{O} (\varepsilon'^{\, 8}, \alpha'^{\, 3}) & ( r > R ) , \\
  \displaystyle 
  \mathcal{D}_3 r^3 + \dfrac{\mathcal{D}_4}{r^4} - 32 \sqrt{\dfrac{3 \pi}{7}} \zeta \chi \left(\dfrac{M}{R}\right)^2 \left(\dfrac{r}{R}\right) J_2 \bigg\{1 - \dfrac{J_2}{2} \left(1 + \dfrac{7446}{2401} \dfrac{r^2}{R^2}
    \right.
    \\
    \left.
      - \dfrac{12186}{343} \dfrac{r^2}{R^2} \ln  r + \dfrac{4740}{343} \dfrac{r^2}{R^2} \ln  \dfrac{r}{R} - \dfrac{2370}{49} \dfrac{r^2}{R^2} \left(\ln  \dfrac{r}{R}\right)^2  \right)  \bigg\}  + \mathcal{O} (\varepsilon'^{\, 8}, \alpha'^{\, 3}) & ( r \leq R ) ,
  \end{cases}
  \\
  \label{eq:SolH5}
&  H_5^{-1} = 
  \begin{cases}
  \displaystyle 
   \mathcal{C}_5 r^5 + \dfrac{\mathcal{C}_6}{r^6} + \dfrac{544}{343} \sqrt{\dfrac{30 \pi}{11}} \zeta \chi \left(\dfrac{M}{r}\right)^2 \left(\dfrac{R}{r} \right)^7 (J_2)^2 \left(1 + \dfrac{392}{289} \dfrac{R^3}{r^3} \right)+ \mathcal{O} (\varepsilon'^{\, 8}, \alpha'^{\, 3}) & ( r > R ) , \\
  \displaystyle 
  \mathcal{D}_5 r^5 + \dfrac{\mathcal{D}_6}{r^6} - \dfrac{18976}{441} \sqrt{\dfrac{10 \pi}{33}} \zeta \chi \left(\dfrac{M}{R} \right)^2 \left(\dfrac{r}{R}\right)^3 (J_2)^2 \left(1  + \dfrac{1710}{593} \ln \dfrac{r}{R} \right)+ \mathcal{O} (\varepsilon'^{\, 8}, \alpha'^{\, 3}) & ( r \leq R ) ,
  \end{cases}
\end{align}
up to ${\cal{O}}(J_2^2)$, where $\mathcal{C}_i$ and $\mathcal{D}_i$ ($i = 1, 2, 3, 4, 5, 6$) are integration constants. 

Let us now determine these constants of integration. Asymptotic flatness and regularity at the center imply that $\mathcal{C}_1$,  $\mathcal{C}_3$, $\mathcal{C}_5$, $\mathcal{D}_2$, $\mathcal{D}_4$ and $\mathcal{D}_6$ must all vanish, and thus, we need two more boundary conditions at the surface to determine the remaining constants. Requiring the matching condition at the surface, we have
\begin{align}
  \lim_{\lambda \to 0^{+}} \left[ H_{\rm ext} ( R + \lambda ) - H_{\rm int} ( R - \lambda ) \right] = &0 ,  \\
  \lim_{\lambda \to 0^{+}} \left[ \left. \frac{d H_{\rm ext}}{d r} \right|_{r = R + \lambda} - \left. \frac{d H_{\rm int}}{d r} \right|_{r = R - \lambda} \right]
  =&-24 \zeta \chi \frac{M^2}{R^3}  \bigg[
  1- \frac{2509 J_2}{588} \bigg\{
  1+ \frac{5619}{2509} \cos2\theta \nonumber\\ &- \frac{8750 J_2}{2509} \bigg(1+ \frac{69}{50} \cos2\theta +\frac{243}{350} \cos4\theta\bigg)
\bigg\} \bigg] \sin\theta e^{-i\phi},
\end{align}
up to ${\cal{O}}(J_2^2)$. 

With this, we finally obtain 
\begin{align}
{}
H_{\rm ext} =& 8 \zeta \chi \bigg( \dfrac{M}{r} \bigg)^2 \bigg( \dfrac{R}{r} \bigg)^3 \bigg[1 - \dfrac{1}{3} \dfrac{R^3}{r^3} -J_2 \bigg\{1 + \dfrac{261}{3430} \dfrac{r}{R} + \dfrac{9}{98} \dfrac{R^2}{r^2} + \dfrac{567}{572} \dfrac{R^5}{r^5} \nonumber\\
& + \dfrac{87}{686} \dfrac{r}{R} \bigg[1 + \dfrac{721}{29} \dfrac{R^3}{r^3} + \dfrac{1715}{754} \dfrac{R^6}{r^6} \bigg] \cos 2 \theta - \dfrac{153 J_2}{4802} \dfrac{r}{R} \bigg(1 - \dfrac{25165}{1836} \dfrac{R^2}{r^2} \nonumber\\
& - \dfrac{2401}{34} \dfrac{R^3}{r^3} + \dfrac{833}{8} \dfrac{R^5}{r^5} + \dfrac{765919}{7514} \dfrac{R^8}{r^8} + \dfrac{5}{3} \bigg[1 + \dfrac{35231}{918} \dfrac{R^2}{r^2} - \dfrac{2401}{170} \dfrac{R^3}{r^3} \nonumber\\
&+ \dfrac{147}{4} \dfrac{R^5}{r^5} + \dfrac{50421}{578} \dfrac{R^8}{r^8} + \dfrac{35231}{612} \dfrac{R^2}{r^2} \bigg(1 + \dfrac{459}{1438} \dfrac{R^3}{r^3} + \dfrac{5292}{12223} \dfrac{R^6}{r^6}  \bigg) \cos 2\theta  \bigg] \cos 2\theta  \bigg) \bigg\}  \bigg] \sin \theta \; e^{-i \phi} \nonumber\\
&+ \mathcal{O} (\varepsilon'^{\, 8}, \alpha'^{\, 3}) , \\
  H_{\rm int} = & \dfrac{16}{3} \zeta \chi \bigg( \dfrac{M}{R} \bigg)^2  \bigg( \dfrac{r}{R} \bigg) \bigg[1 - \dfrac{1497 J_2}{440} \bigg\{1 - \dfrac{106920}{2225041} \dfrac{r^2}{R^2} - \dfrac{1100}{499} \ln \dfrac{r}{R}  + \dfrac{825}{499} \bigg[1 - \dfrac{216}{4459} \dfrac{r^2}{R^2}  \bigg] \cos 2\theta \nonumber\\
  &+ \dfrac{3666795 J_2}{4450082} \bigg(1 - \dfrac{11694007}{2800098} \dfrac{r^2}{R^2}  + \dfrac{409500}{377791} \dfrac{r^4}{R^4}   - \dfrac{22295}{22223} \bigg[1 + \dfrac{600}{119} \dfrac{r^4}{R^4} - \dfrac{588424}{324135} \dfrac{r^2}{R^2} \nonumber\\
  &- \dfrac{593}{126} \dfrac{r^2}{R^2} \bigg(1 - \dfrac{16200}{10081} \dfrac{r^2}{R^2} \bigg) \cos 2\theta  \bigg] \cos 2\theta + \dfrac{89180}{66669} \ln \dfrac{r}{R}  \bigg\{1 + \dfrac{25167}{1372} \dfrac{r^2}{R^2} \nonumber\\
  &+ \dfrac{3645}{98} \dfrac{r^2}{R^2} \ln \dfrac{r}{R} + \dfrac{7912}{343} \dfrac{r^2}{R^2} \bigg[1 + \dfrac{24885}{15824} \ln \dfrac{r}{R} + \dfrac{13965}{31648} \cos 2\theta  \bigg] \cos 2\theta  \bigg\} \bigg)  \bigg\}  \bigg] \sin \theta \; e^{-i \phi} \nonumber\\
  & + \mathcal{O} (\varepsilon'^{\, 8}, \alpha'^{\, 3}) , 
\end{align}
up to ${\cal{O}}(J_2^2)$. By using the definition of $H$, we then find the gravitomagnetic sector for dCS metric deformation up to ${\cal{O}}(J_2^2)$
\begin{align}
  \delta h^{\rm ext}_{0 x} = & - 8 \zeta \chi \bigg( \dfrac{M}{r} \bigg)^2 \bigg( \dfrac{R}{r} \bigg)^3 \bigg[1 - \dfrac{1}{3} \dfrac{R^3}{r^3} -J_2 \bigg\{1 + \dfrac{261}{3430} \dfrac{r}{R} + \dfrac{9}{98} \dfrac{R^2}{r^2} + \dfrac{567}{572} \dfrac{R^5}{r^5} \nonumber\\
& + \dfrac{87}{686} \dfrac{r}{R} \bigg[1 + \dfrac{721}{29} \dfrac{R^3}{r^3} + \dfrac{1715}{754} \dfrac{R^6}{r^6} \bigg] \cos 2 \theta - \dfrac{153 J_2}{4802} \dfrac{r}{R} \bigg(1 - \dfrac{25165}{1836} \dfrac{R^2}{r^2} \nonumber\\
& - \dfrac{2401}{34} \dfrac{R^3}{r^3} + \dfrac{833}{8} \dfrac{R^5}{r^5} + \dfrac{765919}{7514} \dfrac{R^8}{r^8} + \dfrac{5}{3} \bigg[1 + \dfrac{35231}{918} \dfrac{R^2}{r^2} - \dfrac{2401}{170} \dfrac{R^3}{r^3} \nonumber\\
&+ \dfrac{147}{4} \dfrac{R^5}{r^5} + \dfrac{50421}{578} \dfrac{R^8}{r^8} + \dfrac{35231}{612} \dfrac{R^2}{r^2} \bigg(1 + \dfrac{459}{1438} \dfrac{R^3}{r^3} + \dfrac{5292}{12223} \dfrac{R^6}{r^6}  \bigg) \cos 2\theta  \bigg] \cos 2\theta  \bigg) \bigg\}  \bigg] \sin \theta \;  \sin \phi \nonumber\\
&+ \mathcal{O} (\varepsilon'^{\, 8}, \alpha'^{\, 3}) , \\
  \delta h^{\rm ext}_{0 y} = & 8 \zeta \chi \bigg( \dfrac{M}{r} \bigg)^2 \bigg( \dfrac{R}{r} \bigg)^3 \bigg[1 - \dfrac{1}{3} \dfrac{R^3}{r^3} -J_2 \bigg\{1 + \dfrac{261}{3430} \dfrac{r}{R} + \dfrac{9}{98} \dfrac{R^2}{r^2} + \dfrac{567}{572} \dfrac{R^5}{r^5} \nonumber\\
& + \dfrac{87}{686} \dfrac{r}{R} \bigg[1 + \dfrac{721}{29} \dfrac{R^3}{r^3} + \dfrac{1715}{754} \dfrac{R^6}{r^6} \bigg] \cos 2 \theta - \dfrac{153 J_2}{4802} \dfrac{r}{R} \bigg(1 - \dfrac{25165}{1836} \dfrac{R^2}{r^2} \nonumber\\
& - \dfrac{2401}{34} \dfrac{R^3}{r^3} + \dfrac{833}{8} \dfrac{R^5}{r^5} + \dfrac{765919}{7514} \dfrac{R^8}{r^8} + \dfrac{5}{3} \bigg[1 + \dfrac{35231}{918} \dfrac{R^2}{r^2} - \dfrac{2401}{170} \dfrac{R^3}{r^3} \nonumber\\
&+ \dfrac{147}{4} \dfrac{R^5}{r^5} + \dfrac{50421}{578} \dfrac{R^8}{r^8} + \dfrac{35231}{612} \dfrac{R^2}{r^2} \bigg(1 + \dfrac{459}{1438} \dfrac{R^3}{r^3} + \dfrac{5292}{12223} \dfrac{R^6}{r^6}  \bigg) \cos 2\theta  \bigg] \cos 2\theta  \bigg) \bigg\}  \bigg] \sin \theta \;  \cos \phi \nonumber\\
& + \mathcal{O} (\varepsilon'^{\, 8}, \alpha'^{\, 3}) , \\
 \delta h^{\rm int}_{0 x} = & - \dfrac{16}{3} \zeta \chi \bigg( \dfrac{M}{R} \bigg)^2  \bigg( \dfrac{r}{R} \bigg) \bigg[1 - \dfrac{1497 J_2}{440} \bigg\{1 - \dfrac{106920}{2225041} \dfrac{r^2}{R^2} - \dfrac{1100}{499} \ln \dfrac{r}{R}  + \dfrac{825}{499} \bigg[1 - \dfrac{216}{4459} \dfrac{r^2}{R^2}  \bigg] \cos 2\theta \nonumber\\
  &+ \dfrac{3666795 J_2}{4450082} \bigg(1 - \dfrac{11694007}{2800098} \dfrac{r^2}{R^2}  + \dfrac{409500}{377791} \dfrac{r^4}{R^4}   - \dfrac{22295}{22223} \bigg[1 + \dfrac{600}{119} \dfrac{r^4}{R^4} - \dfrac{588424}{324135} \dfrac{r^2}{R^2} \nonumber\\
  &- \dfrac{593}{126} \dfrac{r^2}{R^2} \bigg(1 - \dfrac{16200}{10081} \dfrac{r^2}{R^2} \bigg) \cos 2\theta  \bigg] \cos 2\theta + \dfrac{89180}{66669} \ln \dfrac{r}{R}  \bigg\{1 + \dfrac{25167}{1372} \dfrac{r^2}{R^2} \nonumber\\
  &+ \dfrac{3645}{98} \dfrac{r^2}{R^2} \ln \dfrac{r}{R} + \dfrac{7912}{343} \dfrac{r^2}{R^2} \bigg[1 + \dfrac{24885}{15824} \ln \dfrac{r}{R} + \dfrac{13965}{31648} \cos 2\theta  \bigg] \cos 2\theta  \bigg\} \bigg)  \bigg\}  \bigg] \sin \theta \; \sin\phi \nonumber\\
  & + \mathcal{O} (\varepsilon'^{\, 8}, \alpha'^{\, 3}) , \\
   \delta h^{\rm int}_{0 y} = & \dfrac{16}{3} \zeta \chi \bigg( \dfrac{M}{R} \bigg)^2  \bigg( \dfrac{r}{R} \bigg) \bigg[1 - \dfrac{1497 J_2}{440} \bigg\{1 - \dfrac{106920}{2225041} \dfrac{r^2}{R^2} - \dfrac{1100}{499} \ln \dfrac{r}{R}  + \dfrac{825}{499} \bigg[1 - \dfrac{216}{4459} \dfrac{r^2}{R^2}  \bigg] \cos 2\theta \nonumber\\
  &+ \dfrac{3666795 J_2}{4450082} \bigg(1 - \dfrac{11694007}{2800098} \dfrac{r^2}{R^2}  + \dfrac{409500}{377791} \dfrac{r^4}{R^4}   - \dfrac{22295}{22223} \bigg[1 + \dfrac{600}{119} \dfrac{r^4}{R^4} - \dfrac{588424}{324135} \dfrac{r^2}{R^2} \nonumber\\
  &- \dfrac{593}{126} \dfrac{r^2}{R^2} \bigg(1 - \dfrac{16200}{10081} \dfrac{r^2}{R^2} \bigg) \cos 2\theta  \bigg] \cos 2\theta + \dfrac{89180}{66669} \ln \dfrac{r}{R}  \bigg\{1 + \dfrac{25167}{1372} \dfrac{r^2}{R^2} \nonumber\\
  &+ \dfrac{3645}{98} \dfrac{r^2}{R^2} \ln \dfrac{r}{R} + \dfrac{7912}{343} \dfrac{r^2}{R^2} \bigg[1 + \dfrac{24885}{15824} \ln \dfrac{r}{R} + \dfrac{13965}{31648} \cos 2\theta  \bigg] \cos 2\theta  \bigg\} \bigg)  \bigg\}  \bigg] \sin \theta \; \cos\phi \nonumber\\
  & + \mathcal{O} (\varepsilon'^{\, 8}, \alpha'^{\, 3}).
\end{align}

%%%%%%%%%%%%%%%%%%%%%%%%%%%%%%%%%%%%%%%%%%%%%%%%% 
\section{The gyroscope coordinate system}
\label{App:GyroCoord}

In the GPB experiment, the spin vector is along the direction of IM Pegasi~\cite{Everitt:2011hp}. For such a Solar System experiment with a satellite orbiting around Earth, it is convenient to use inertial coordinates defined by the spin vector $\bm{S}$ and the spin-angular momentum vector of Earth $\bm{J}_{\rm E}$. We choose the $x'$-axis, $y'$-axis, and $z'$-axis as the directions of $\bm{S} \times \bm{J}_{\rm E}$, $\bm{S}$, and $(\bm{S} \times \bm{J}_{\rm E}) \times \bm{S}$, respectively. Therefore, from Eq.~\eqref{spin}, precession in the $y'$-direction always vanishes, and thus, we only investigate precession in the $x'$- and $z'$-directions. Let us call such a set of coordinates $(x', y', z')$ the {\it gyroscope coordinate system}. The $x'$-axis and the $z'$-axis correspond to the west-east (WE) and the north-south (NS) directions in~\cite{Everitt:2011hp}, respectively.

In this coordinate system, the components of $\bm{J}_{\rm E}$ and $\hat{\bm{S}}$ are
\begin{align}
\bm{J}_{\rm E} = J_{\rm E} 
\begin{pmatrix}
0 \\
\sin I_S \\
\cos I_S 
\end{pmatrix} ,
\qquad
\hat{\bm{S}} = 
\begin{pmatrix}
0 \\
1 \\
0
\end{pmatrix} ,
\end{align}
where $I_S$ is the declination of the guide star (e.g. the IM Pegasi in the GPB experiment). For simplicity, let us also assume the satellite is in circular motion around Earth. In this case, the components of the unit vectors $\bm{n}_{\rm gyro}$ and $\bm{h}_{\rm gyro}$ can be expressed in the gyroscope coordinate system as
\begin{align}
\bm{n}_{\rm gyro} &= \sin ( \omega_{\rm gyro} t + \varphi_{\rm gyro} ) \, \bm{n}^{s}_{\rm gyro} + \cos ( \omega_{\rm gyro} t + \varphi_{\rm gyro} ) \, \bm{n}^{c}_{\rm gyro} , \\
\bm{h}_{\rm gyro} &= 
\begin{pmatrix}
\sin I_{\rm gyro} \cos ( \varphi_S - \Omega_{\rm gyro} )
 \\
\sin I_S \cos I_{\rm gyro} 
- \cos I_S \sin I_{\rm gyro} \sin ( \varphi_S - \Omega_{\rm gyro} )
 \\
\cos I_S \cos I_{\rm gyro} 
+ \sin I_S \sin I_{\rm gyro} \sin ( \varphi_S - \Omega_{\rm gyro} )
\end{pmatrix} , 
\end{align}
where we have defined
\begin{align}
  \bm{n}^{s}_{\rm gyro} =
  & 
    \begin{pmatrix}
      - \cos I_{\rm gyro} \cos ( \varphi_S - \Omega_{\rm gyro} ) \\ 
      [ \sin I_S \sin I_{\rm gyro} + \cos I_S \cos I_{\rm gyro} \sin ( \varphi_S - \Omega_{\rm gyro} ) ] \\
      [ \cos I_S \sin I_{\rm gyro} - \sin I_S \cos I_{\rm gyro} \sin ( \varphi_S - \Omega_{\rm gyro} ) ]
    \end{pmatrix} ,
  \\
  \bm{n}^{c}_{\rm gyro} =
  &
    \begin{pmatrix}
      \sin ( \varphi_S - \Omega_{\rm gyro} ) \\ 
      \cos I_S \cos ( \varphi_S - \Omega_{\rm gyro} ) \\
      - \sin I_S \cos ( \varphi_S - \Omega_{\rm gyro} ) 
    \end{pmatrix} ,
\end{align}
$\omega_{\rm gyro}$ is the orbital angular velocity of the satellite around Earth, $I_{\rm gyro}$, $\varphi_{\rm gyro}$, and $\Omega_{\rm gyro}$ are the inclination, initial direction, and right ascension of the ascending node of the satellite, respectively, and $\varphi_S$ is the right ascension of the guide star.

Assuming the rotating object is Earth, the precession rate [Eq.~\eqref{Eq:dS_average}] for the Lense-Thirring effect in GR becomes
\begin{align}
\left\langle \bm{P}_r \right\rangle_{T_{\rm gyro}}^{\rm LT} = 
\frac12 \frac{1}{r_{\rm gyro}^3} 
\left[ \bm{J}_{\rm E} 
- 3 ( \bm{J}_{\rm E} \cdot \bm{h}_{\rm gyro} ) \bm{h}_{\rm gyro} \right] 
\times \hat{\bm{S}} .
\end{align}
Comparing this with the effect from a dCS metric deformation, as given in Eq.~\eqref{Eq:Pr_dCS}, one can see that both effects contribute in the same direction.  If the satellite is launched in a polar orbit, i.e. $I_{\rm gyro} = \pi/2$, then we obtain the following WE and North-South (NS) components for the Lense-Thirring precession rate 
\begin{align}
  {}
  \left[ \left\langle \bm{P}_r \right\rangle_{T_{\rm gyro}}^{\rm LT} \right]_{\rm WE} &= - \frac12 \frac{G |\bm{J}_{\rm E}|}{c^2 r_{\rm gyro}^3} \cos I_S , \\
  \left[ \left\langle \bm{P}_r \right\rangle_{T_{\rm gyro}}^{\rm LT} \right]_{\rm NS} &= 0 . &
\end{align}
Therefore, especially in the case of the GPB experiment, where the spin vector is along the direction of the IM Pegasi, we recover the well-known result
\begin{align}
\left[ \left\langle \bm{\delta}_{\rm LT} \right\rangle_{P_{\rm gyro}} \right]_{\rm WE}
= - 39 \, [{\rm mas/yr}] .
\end{align}

%%%%%%%%%%%%%%%%%%%%%%%%%%%%%%%%%%%%%%%%%%%%%%%%%%%%%%%%
%%%%%%%%%%%%%%%%%%%%%%%%%%%%%%%%%%%%%%%%%%%%%%%%%%%%%%%%

\bibliography{CSpaper-N-blx}

\end{document}